\documentclass[fleqn,10pt]{wlscirep}
\usepackage{mathrsfs}
\usepackage{braket}
\usepackage{graphicx}
\usepackage{amsfonts}
\usepackage{enumitem}
\usepackage{etoolbox}
\usepackage{threeparttable, tablefootnote}
\usepackage[numbers,sort&compress]{natbib}
\usepackage[final]{pdfpages}

\title{Analysis of Side-band Inequivalence}

\author[1,*]{Sina Khorasani}
\affil[1]{Vienna Center for Quantum Science and Technology, Boltzmanngasse 5, 1090 Vienna, Austria}
\affil[*]{sina.khorasani@ieee.org}

\begin{document}

\begin{abstract}
Frequency shifts of red- and blue-scattered (Stokes/anti-Stokes) side-bands in quantum optomechanics are shown to be counter-intuitively inequal, resulting in an unexpected symmetry breaking. This difference is referred to as Side-band Inequivalenve (SI), which normally leans towards red, and being a nonlinear effect it depends on optical power or intracavity photon number. Also there exists a maximum attainable SI at an optimal operation point. The mathematical method employed here is a combination of operator algebra equipped with harmonic balance, which allows a clear understanding of the associated nonlinear process. This reveals the existence of three distinct operation regimes in terms of pump power, two of which have immeasurably small SI. Compelling evidence from various experiments sharing similar interaction Hamiltonians, including quantum optomechanics, ion/Paul traps, electrooptic modulation, Brillouin scattering, and Raman scattering unambiguously confirm existence of a previously unnoticed SI.
\end{abstract}

	\flushbottom
\maketitle

\thispagestyle{empty}


The nonlinearity of optomechanical interaction \cite{0a,0b,0c,0d,0e,0f,0g,0h,0i,0j} causes scattering of incident photons with the annihilator $\hat{a}$ from the cavity unto either red- or blue-shifted photons through annihilation or generation of a cavity phonon with the annihilator $\hat{b}$, giving rise to the first-order mechanical side-bands. Taking the optical frequency $\omega$ to be at a detuning $\Delta=\omega_{\rm c}-\omega$ from cavity resonance $\omega_{\rm c}$, the $\nu-$th order sidebands are naturally expected to occur at the detunings $\Delta_{\pm\nu}=\Delta\mp \nu\Omega$, where $\Omega$ represents the mechanical frequency. As a results, the first-order mechanical side-bands of scattered red $\Delta_{+1}$ and blue $\Delta_{-1}$ processes must average out back to the original pump detuning $\Delta$. 

\begin{figure}[ht!]
	\centering
	\includegraphics[width=5.98in]{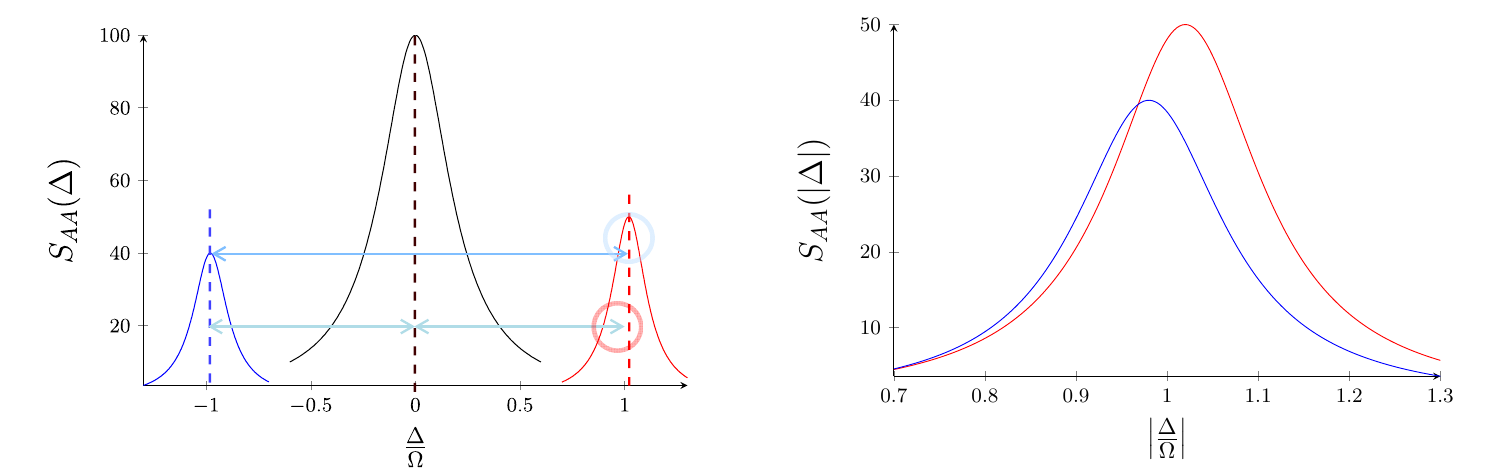}
	\caption{Illustration of two basic symmetry breakings in a hypothetical quantum optomechanical cavity at resonant pump, and in absence of cooling tone: (left) separated side-bands in a heterodyne setup; (right) zoom out of overlapping side-bands in a homodyne detection around mechanical frequency with $|\Delta|=\Omega$. The red detuning exceeds that of blue detuning, shown in pale red circle, corresponding to the side-band inequivalence (pale red circle), and also the height of red peak exceeds that of blue peak, corresponding to side-band asymmetry (pale blue circle). Here, $S_{\rm AA}(\Delta)$ represents the spectral noise density. Blue and red Lorentzian linewidths are $0.1\Omega$, while the central resonance has a linewidth of $0.2\Omega$. Side-band Inequivalence is chosen to be $\bar{\delta}=\delta/\Omega=0.02$ towards red side-band. \label{Figure1}}
\end{figure}

Defining the deviation of this average from $\Delta$, as $\delta=\frac{1}{2}(\Delta_{+1}+\Delta_{-1})-\Delta$, then one may normally expect $\delta=0$. A non-zero $\delta$ would have otherwise implied the so-called Side-band Inequivalence (SI). This type of asymmetry appears to have a classical nonlinear nature, and is illustrated in Fig. \ref{Figure1}.

There is, however, another well-known type of side-band asymmetry between the red and blue side-bands in the context of optomechanics, which has a quantum nature and may be used for instance to accurately determine the absolute temperature through a reference-free optomechanical measurement \cite{Purdy1,Purdy2}. This is based on the ratio of Stokes to anti-Stokes Raman transition rates, which is equal to $\exp(\hbar\Omega/k_{\rm B}T)$ where $k_{\rm B}$ is the Boltzmann's constant and $T$ is the absolute temperature \cite{asym1,asym2}. Also illustrated in Fig. \ref{Figure1}, clearly, SI is quite different from this type of side-band asymmetry.

While both time-reversal symmetry and energy conservation are fundamentally preserved in this scattering process, a nonlinear analysis of quantum optomechanics using the recently developed method of higher-order operators \cite{1,1a0,1a,1b,1c} necessitates a slight difference among detunings of blue and red-scattered photons, the amount of which was initially found to increase roughly proportional to the intracavity photon number $\bar{n}$. Here, $\bar{n}$ is defined as the steady-state mean-value of the number operator $\hat{n}=\hat{a}^\dagger\hat{a}$. 

Surprisingly enough, this disagreement satisfying $\delta\neq 0$ does not violate the energy conservation law, actually allowed by the finite cavity linewidth as well as the single-photon/single-phonon nature of the process involved. Moreover, the time-reversal symmetry is also preserved.

Among the pool of available experimental data, only a handful of side-band resolved cavities reveal this disagreement \cite{1}. Some initial trial experiments recently done at extremely high intracavity photon numbers $\bar{n}$, and/or extremely large single-photon optomechanical interaction rates $g_0$, though, failed to demonstrate its existence. This may raise the speculation that whether SI would have been merely a mathematical artifact, or something has been missing due to not doing the operator analysis to the highest-order.

A careful analysis of this phenomenon, however, confirms the latter, thus classifying the quantum optomechanical interaction into three distinct regimes with different behaviors:
\begin{itemize}[leftmargin=*]
	\setlength\itemsep{-0.4em}
	\item Fully Linear: This regime can be investigated using the lowest-order analysis and first-order operators, which is conventionally done by linearizing the Hamiltonian around equilibrium points. This will require the four-dimensional basis of first-order ladder operators $\{\hat{a},\hat{a}^\dagger,\hat{b},\hat{b}^\dagger\}$ and is indeed quite sufficient to understand many of the complex quantum optomechanical phenomena \cite{2}.
	\item Weakly Nonlinear: This regime requires higher-order operator analysis of at least second-order. This can be done using the three-dimensional reduced basis \cite{1,2} given as $\{\hat{a},\hat{a}\hat{b},\hat{a}\hat{b}^\dagger\}$.
	\item Strongly Nonlinear: Full understanding of this interaction regime requires the highest-order analysis using third-order operators. Referred as to the minimal basis \cite{1,2}, the convenient reduced choice is the two-dimensional basis $\{\hat{n}^2,\hat{n}\hat{b}\}$. 
\end{itemize}

In the latter regime, the behavior of governing equations is in such a way that a fully-linearized analysis of fluctuations mostly happens to work. This definitely sounds strange, unless we note that for the strongly non-linear regime, the optical pump level is so strong that the quantumness of intracavity photons disappear and we can do the replacements $\hat{n}(t)\rightarrow\bar{n}(t)$ and $\hat{a}(t)\rightarrow e^{i\Delta t}{\rm a}(t)$. Once the steady-state is reached, the minimal basis reduces to ${\bar{n}^2,\bar{n}\hat{b}}$. It is easy to see that $\bar{n}$ stays almost constant so that mechanics is still quantized and expressed via $\hat{b}$. Based on the Langevin's equations for classical electromagnetic field ${\rm a}(t)$ and quantized mechanics $\hat{b}(t)$ it is easy to see that the only surviving frequencies are $\Delta\pm\Omega\pm2\Omega\pm\cdots$. Hence, quantumness of optics is absolutely necessary for the SI to occur, and this disappears when the optical pump is so strong that the electromagnetic field can be treated classically.

Side-band Inequivalence is essentially forbidden in the fully linear regime, and it also quickly fades away in the strongly nonlinear regime. But it may only happen in the weakly nonlinear regime. This is now also confirmed both by the higher-order operator method and extensive calculations. It typically does not exceed one part in million to one part in ten thousand in solid state quantum optomechanics, and therefore, it is a very delicate phenomenon and elusive to observe. As it will be shown, other experiments such as Raman scattering may instead exhibit much stronger inequivalences.

This letter provides a direct route towards clear understanding of this complex nonlinear phenomenon. Using a combination of operator algebra and harmonic balance (used in analysis of laser diodes) \cite{Cabon}, we obtain a closed form expression for SI $\delta$ as a function of intracavity photon number $\bar{n}$, which is expected to be valid through all three above operation regimes, and for any arbitrarily chosen set of optomechanical parameters. Not only the findings of this work reproduce the approximate linear expression found earlier through second-order operators \cite{1} for the weakly nonlinear regime, but also we can show that there is an optimal point at which the SI attains a maximum. Moving away from the optimal point, both at the much smaller and much larger pump rates, $\delta$ attains much smaller values, tending to zero in the limit of very large $\bar{n}$. An analytical treatment of nonlinear optomechanical equations based on breathing solutions is also presented, which again confirms the existence of inequivalence towards red and is presented in the Supplementary Information.

This will be greatly helpful to designate the investigation range of experimental parameters given any available optomechanical cavity. Furthermore, it marks a clear and definable border among the three above-mentioned operation regimes. 

\section*{Results}

The analysis of SI proceeds with considering the behavior of optomechanical cavity under steady-state conditions. We will focus only on the first-order side-bands and discard all other contributions coming from or to the second- and higher-order side-bands. We consider a single-frequency pump with ideally zero linewidth at a given detuning $\Delta$, which normally gives rise to two stable blue- and red- side-bands. Hence, the time-dependence of the photon annihilator will look like
\begin{equation}
\label{eq1}
\hat{a}(t)=\hat{a}_0 e^{i\Delta t}+\hat{a}_b e^{i(\Delta-\Omega+\frac{1}{2}\delta)t}+\hat{a}_r e^{i(\Delta+\Omega+\frac{1}{2}\delta)t}+\cdots,
\end{equation}
\noindent
where $\hat{a}_0$, $\hat{a}_b$, and $\hat{a}_r$ respectively correspond to the central excitation resonance at pump frequency, and blue- and red-detuned side-bands. In the weak coupling regime where $g=g_0\sqrt{\bar{n}}<<\Omega$, the $n+1-$th order side-band is typically $(g/\Omega)^n$ times weaker in amplitude than the 1st-order side-band. This means, for instance, that the second-order side-band is $g/\Omega$ weaker than the first-order side-bands. In that sense, their respective contributions and significance diminish very rapidly with their growing order. So, there is good reason to agree that the truncation in (\ref{eq1}) up to the first-order side-bands is a rather acceptable approximation. Inclusion of 2nd-order terms such as $\hat{a}_{bb}e^{i(\Delta-2\Omega+\delta)t}$ and $\hat{a}_{rr}e^{i(\Delta+2\Omega+\delta)t}$, will result in minor modification of following expressions. It has to be noticed that the red and blue time dependences of $n-$th order processes must be $e^{i(\Delta-n\Omega+\frac{n}{2}\delta)t}$ and $e^{i(\Delta-n\Omega+\frac{n}{2}\delta)t}$, showing that the associated SI with the $n-$th order side-bands should be $n\delta$.

We here furthermore have ignored the symmetrical movement of side-bands which should correspond to the optomechanical spring effect $\delta\Omega$ \cite{0a,0c}, but this correction is zero at $\Delta=0$ while $\delta$ is independent of $\Delta$ (in Supplementary Information we show that contribution of this term is insignificant). It is the asymmetric movements of side-bands which gives rise to SI. In order to do so, it would have been enough to take the ansatz $\hat{a}(t)=\hat{a}_0 e^{i\Delta t}+\hat{a}_b e^{i(\Delta-\varpi+\frac{1}{2}\delta)t}+\hat{a}_r e^{i(\Delta+\varpi+\frac{1}{2}\delta)t}+\cdots$ instead, with $\varpi=\Omega+\delta\Omega$ being the shifted mechanical frequency due to the optomechanical spring effect.

The steady-state time-average of central excitation satisfies $\braket{\hat{a}_0}=\sqrt{\bar{n}}$, where $\bar{n}$ can be determined by solution of a third-order algebraic equation once the optical power pump rate $P_{\rm op}$, detuning $\Delta$, external coupling $\eta$ and all other optomechanical parameters are known. The standard set of basic optomechanical parameters needed here are mechanical frequency $\Omega$, optical decay rate $\kappa$, and mechanical decay rate $\Gamma$. Therefore, the photon number operator up to the first side-bands will behave as
\begin{eqnarray}
\label{eq2}
\hat{n}(t)&=&\hat{a}_0^\dagger\hat{a}_0+\hat{a}_b^\dagger\hat{a}_b+\hat{a}_r^\dagger\hat{a}_r\\ \nonumber
&+&\hat{a}_b^\dagger\hat{a}_0 e^{-i(-\Omega+\frac{1}{2}\delta)t}+\hat{a}_0^\dagger\hat{a}_b e^{i(-\Omega+\frac{1}{2}\delta)t}+\hat{a}_r^\dagger\hat{a}_0 e^{-i(\Omega+\frac{1}{2}\delta)t}+\hat{a}_0^\dagger\hat{a}_r e^{i(\Omega+\frac{1}{2}\delta)t}+\cdots,
\end{eqnarray}
\noindent
while the mechanical annihilator will exhibit a closely spaced doublet around the mechanical frequency spaced within $\delta$ as
\begin{equation}
\label{eq3}
\hat{b}(t)=\hat{b}_0+\hat{b}_b e^{-i(\Omega-\frac{1}{2}\delta)t}+\hat{b}_r e^{-i(\Omega+\frac{1}{2}\delta)t}+\cdots .
\end{equation}
Inclusion of 2nd-order side-bands would have generated extra terms in (\ref{eq2}) such as $\hat{a}_{bb}^\dagger\hat{a}_{r}$ and $\hat{a}_{rr}^\dagger\hat{a}_{b}$ and their conjugates as well as $\hat{a}_{bb}^\dagger\hat{a}_{bb}+\hat{a}_{rr}^\dagger\hat{a}_{rr}$. For instance, $\hat{a}_{bb}^\dagger\hat{a}_{r}$ should have added up to $\hat{a}_{b}^\dagger\hat{a}_{0}$ and so on. These ignored terms all would cause third- or fourth-order corrections to (\ref{eq2}), which obviously were dropped.

Here, the average mechanical displacement satisfies
\begin{equation}
\label{eq4}
b_0=\braket{\hat{b}_0}=\frac{i g_0 \bar{n}}{i\Omega+\frac{1}{2}\Gamma}.
\end{equation} 
Now, let us get back to the Langevin equation for mechanical motions, which simply is
\begin{equation}
\frac{d}{dt}\hat{b}(t)=(-i\Omega-\frac{1}{2}\Gamma)\hat{b}(t)-i g_0\hat{n}(t)+\sqrt{\Gamma}\hat{b}_{\rm in}(t),
\end{equation}
\noindent
where $\hat{b}_{\rm in}(t)$ is the operator for mechanical fluctuations. For the purpose of our analysis here, all fluctuations can be discarded since they are irrelevant to the formation of side-band frequencies and average out to zero. Using (\ref{eq2}) and (\ref{eq3}) we get
\begin{eqnarray}
\label{eq6}
&&-i(\Omega+\frac{\delta}{2})\hat{b}_r e^{-i(\Omega+\frac{\delta}{2})t}-i(\Omega-\frac{\delta}{2})\hat{b}_b e^{-i(\Omega-\frac{\delta}{2})t}\approx \\ \nonumber
&&-(i\Omega+\frac{\Gamma}{2})\hat{b}_r e^{-i(\Omega+\frac{\delta}{2})t}-(i\Omega+\frac{\Gamma}{2})\hat{b}_b e^{-i(\Omega-\frac{\delta}{2})t}-i g_0 \hat{a}_0^\dagger\hat{a}_b e^{-i(\Omega-\frac{\delta}{2})t}-i g_0 \hat{a}_r^\dagger\hat{a}_0 e^{-i(\Omega+\frac{\delta}{2})t}+\cdots .
\end{eqnarray}
\noindent
From the above, we obtain two key operator equations
\begin{eqnarray}
\label{eq7}
\hat{b}_r&=&\frac{-i 2 g_0}{-i\delta+\Gamma}\hat{a}_r^\dagger\hat{a}_0, \\ \nonumber
\hat{b}_b&=&\frac{-i 2 g_0}{i\delta+\Gamma}\hat{a}_0^\dagger\hat{a}_b.
\end{eqnarray}

In a similar manner, the Langevin equation for the photon annihilator is
\begin{equation}
\label{eq8}
\frac{d}{dt}\hat{a}(t)=(i\Delta-\frac{1}{2}\kappa)\hat{a}(t)+ig_0\hat{a}(t)[\hat{b}(t)+\hat{b}^\dagger(t)]+\sqrt{\kappa}\hat{a}_{\rm in}.
\end{equation}
\noindent
Using (\ref{eq1}) and (\ref{eq3}) we obtain
\begin{eqnarray}
\label{eq9}
&&i\Delta\hat{a}_0 e^{i\Delta t}+i(\Delta-\Omega+\frac{\delta}{2})\hat{a}_b e^{i(\Delta-\Omega+\frac{\delta}{2})t}+i(\Delta+\Omega+\frac{\delta}{2})\hat{a}_r e^{i(\Delta+\Omega+\frac{\delta}{2})t}\approx \\ \nonumber
&&(i\Delta-\frac{\kappa}{2})\left[\hat{a}_0 e^{i\Delta t}+\hat{a}_b e^{i(\Delta-\Omega+\frac{\delta}{2})t}+\hat{a}_r e^{i(\Delta+\Omega+\frac{\delta}{2})t}\right]+ \\ \nonumber
&&i g_0\left[\hat{a}_0 e^{i\Delta t}+\hat{a}_b e^{i(\Delta-\Omega+\frac{\delta}{2})t}+\hat{a}_r e^{i(\Delta+\Omega+\frac{\delta}{2})t}\right]\left[\hat{x}_0+\hat{b}_r e^{-i(\Omega+\frac{\delta}{2})t}+\hat{b}_b e^{-i(\Omega-\frac{\delta}{2})t}+\hat{b}_r^\dagger e^{i(\Omega+\frac{\delta}{2})t}+\hat{b}_b^\dagger e^{i(\Omega-\frac{\delta}{2})t}\right].
\end{eqnarray}
\noindent
where $\hat{x}_0=\hat{b}_0+\hat{b}_0^\dagger$. This will yield the further operator equations as
\begin{eqnarray}
\label{eq10}
\frac{\kappa}{2i g_0}\hat{a}_0&=& \hat{a}_0\hat{x}_0+\hat{a}_b\hat{b}_b^\dagger+\hat{a}_r\hat{b}_r, \\ \nonumber
\left[\frac{i(-\Omega+\frac{\delta}{2})+\frac{\kappa}{2}}{ig_0}\right]\hat{a}_b&=&\hat{a}_b\hat{x}_0+\hat{a}_0\hat{b}_b, \\ \nonumber
\left[\frac{i(\Omega+\frac{\delta}{2})+\frac{\kappa}{2}}{ig_0}\right]\hat{a}_r&=&\hat{a}_r\hat{x}_0+\hat{a}_0\hat{b}_r^\dagger.
\end{eqnarray}


Now, substituting whatever we have in hand in the second equation of (\ref{eq10}), and taking expectation values at the end, we obtain a key algebraic equation in terms of $\delta$ as
\begin{equation}
\label{eq11}
i\left(-\Omega+\frac{1}{2}\delta\right)+\frac{1}{2}\kappa=ig_0 x_0-ig_0\sqrt{\bar{n}}\frac{2ig_0\sqrt{\bar{n}}}{i\delta+\Gamma}.
\end{equation}
\noindent
with $x_0=b_0+b_0^\ast$. Rearrangement of the above gives rise to the equation
\begin{equation}
\label{eq12}
\delta^2-\left[2\Omega+i\gamma+2g_0 x_0\right]\delta+\left[(2i\Omega-\kappa)\Gamma+4g_0^2\bar{n}+2i\Gamma g_0 x_0\right]=0,
\end{equation}
in which $x_0=b_0+b_0^\ast$ and $\gamma=\kappa+\Gamma$ is the total optomechanical decay rate \cite{1,1a}.

This approximate nature of this equation will yield complex values for $\delta$ the imaginary value of which has to be discarded. In practice, we observe that imaginary values are much smaller than real values for practical choices of optomechanical parameters. Furthermore, it leaves room to ignore the square terms $\delta^2$, to admit the solution
\begin{equation}
\label{eq13}
\delta=\Re\left[\frac{(2i\Omega-\kappa)\Gamma+4g_0^2\bar{n}+2i\Gamma g_0 x_0}{2\Omega+i\gamma+2g_0 x_0}\right].
\end{equation}  

This solution can be put into the more convenient form using (\ref{eq4}) and further simplification as
\begin{eqnarray}
\label{eq14}
\delta(\bar{n})&=&\Re\left[\frac{A+B\bar{n}}{C-i D\bar{n}}\right] =\frac{\Re[A C^\ast]+(B\Re[C]-\Im[A]D)\bar{n}}{|C|^2 -2\Im[C]D\bar{n}+D^2 \bar{n}^2}\\ \nonumber
&=&\frac{[2\Gamma^2\Omega]+2\Omega(B-\Gamma D)\bar{n}}{|C|^2 -4\Omega D\bar{n}+D^2 \bar{n}^2},
\end{eqnarray}
where
\begin{eqnarray}
\label{eq15}
A&=&\Gamma(2i\Omega-\kappa)+\frac{4g_0^2\bar{n}(i+1)\Gamma\Omega}{\Omega^2+\frac{1}{4}\Gamma^2}, \\ \nonumber
B&=&\frac{4g_0^2(\Omega-\frac{1}{2}\Gamma)^2}{\Omega^2+\frac{1}{4}\Gamma^2}=B^\ast, \\ \nonumber
C&=&i\gamma+2\Omega, \\ \nonumber
D&=&\frac{4g_0^2\Omega}{\Omega^2+\frac{1}{4}\Gamma^2}=D^\ast.
\end{eqnarray}
\noindent
In a first approximation, $A\approx\Gamma(2i\Omega-\kappa)$ can be used. Further simplifications are shown and discussed in the following.

\subsection*{Limiting Cases}

The expression (\ref{eq14}) obtained for the SI has interesting properties at the limits of zero and infinite intracavity photon number. We may obtain here after some simplification easily the asymptotic expressions
\begin{eqnarray}
\label{eq16}
\lim\limits_{\bar{n}\rightarrow\infty} \delta(\bar{n})&\sim&\frac{\Omega}{\beta\bar{n}}, \\ \nonumber
\lim\limits_{\bar{n}\rightarrow 0} \delta(\bar{n})&\sim&\frac{2\Gamma^2\Omega}{4\Omega^2+\gamma^2}\approx 0,
\end{eqnarray}
where $\beta=2g_0^2/\Omega^2$, while noting that $\Gamma<<\Omega$ and also for a side-band resolved cavity $\kappa<<\Omega$, together which we have $\kappa<\gamma<<\Omega$. 

One should take into account the fact that for Doppler cavities, side-bands normally resolve well enough for a decisive measurement \cite{2}, and the concept of SI is only practically meaningful for side-band resolved cavities. Therefore, the following approximations are valid
\begin{eqnarray}
\label{eq17}
A&\approx&2i\Gamma\Omega, \\ \nonumber
B&\approx&4g_0^2, \\ \nonumber
C&\approx&2\Omega, \\ \nonumber
D&\approx&\frac{4g_0^2}{\Omega},\\ \nonumber
\delta(\bar{n})&\approx&\frac{2\Gamma^2\Omega+8g_0^2\Omega\bar{n}}{\gamma^2+4\Omega^2\left[1-2\left(g_0/\Omega\right)^2\bar{n}\right]^2}. 
\end{eqnarray}

As long as satisfies $\bar{n}<<2|\Im[C]|/D=\Omega^2/g_0^2$, then second order term $\bar{n}^2$ in the denominator of (\ref{eq15}) is negligible and can be ignored. Under this regime, the SI varies almost linearly with $\bar{n}$ as
\begin{eqnarray}
\label{eq18}
\delta(\bar{n})&\approx&\frac{2\Gamma^2\Omega}{4\Omega^2+\gamma^2}+\frac{8g_0^2\Omega(4\Omega^2+\gamma^2+\Gamma^2)}{(\gamma^2+4\Omega^2)^2}\bar{n} \nonumber \\ 
&\approx&\frac{2g_0^2}{\Omega}\bar{n}.
\end{eqnarray}
This result is also well in complete agreement with the expression obtained earlier for the SI \cite{1} in the limit of $g_0<<\Omega$ given as $g_0^2/\Omega$, considering that a factor of $\frac{1}{2}$ must be added as a result of different definition of $\delta$. 

The first immediate conclusion which can be obtained from (\ref{eq18}) is that the SI $\delta$ is always positive, meaning that the detuning frequency of red-sideband should be always a bit larger in magnitude than the blue-sideband. This also agrees with the previous findings of higher-order operator algebra \cite{1}.

Based on (\ref{eq18}), in the weakly-coupling limit of $g_0\sqrt{\bar{n}}<\Gamma$, the normalized SI defined as $\bar{\delta}=\delta/\Omega$ can be simply approximated as a function of mechanical frequency as $\bar{\delta}(\Omega)\approx\frac{1}{2}\Gamma^2/(\Omega^2+\tfrac{1}{4}\gamma^2)\approx\frac{1}{2}\Gamma^2/\Omega^2$. This approximation is fairly convenient in explanation of SI in Raman scattering of various materials which exhibit multiple Raman lines, as shown comprehensively in the Supplementary Information.

\subsection*{Optimum Operation}

The unique mathematical form of (\ref{eq14}) which is composed of a first- and second-order polynomials in terms of $\bar{n}$ respectively, offers a clear maximum at a certain optimum intracavity photon number $\bar{n}_{\rm max}$. To do this, let us first define the dimensionless constants $\alpha=4g_0^2/\Gamma^2$ and $\beta$ already defined under (\ref{eq16}), $\vartheta=\gamma/2\Omega$, and $\psi=\Gamma^2/2\Omega^2$. Then, the SI (\ref{eq17}) can be rewritten as 
\begin{equation}
\label{eq19}
\delta(\bar{n})=\Omega\psi\frac{1+\alpha\bar{n}}{\vartheta^2+(1-\beta\bar{n})^2}.
\end{equation}
This offers the optimum intracavity photon number and thus the maximum attainable SI as
\begin{eqnarray}
\label{eq20}
\bar{n}_{\rm max}&=&\frac{\sqrt{(\alpha+\beta)^2+\alpha^2\vartheta^2}}{\alpha\beta}-\frac{1}{\alpha}\approx\frac{1}{\beta}
=\frac{\Omega^2}{2g_0^2}, \\ \nonumber
\delta_{\rm max}&=&\delta(\bar{n}_{\rm max})\approx\frac{4\Omega^3}{\gamma^2}.
\end{eqnarray}
We should take note of the fact that the maximum practically measureable SI, which occurs at the optimum intracavity photon number $\bar{n}_{\rm max}=\Omega^2/2g_0^2$, is actually at the onset of bistability, and under practical conditions, heating due to optical losses in dielectric. 

\begin{figure}
	\centering
	\includegraphics[width=2.96in]{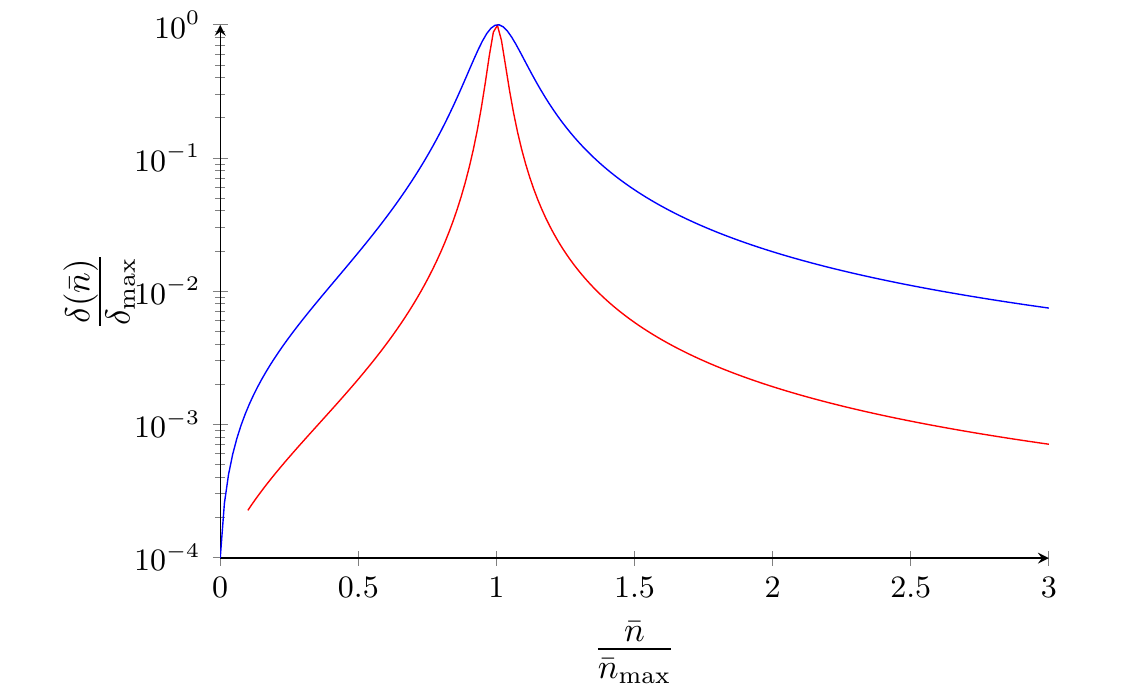}
	\caption{Variation of SI around the maximum point in terms of various settings of parameters: (blue) $\alpha=0.1$, $\beta=10^{-3}$, and $\vartheta=0.1$ (red) $\alpha=0.1$, $\beta=10^{-2}$, and $\vartheta=3.16\times10^{-2}$. Normalization on the vertical axis has been done to drop dependence of $\delta$ on $\Omega\psi$ in (\ref{eq19}). The behavior versus intracavity photon number $\bar{n}$ is clearly not Lorentzian. \label{Fig1}}
\end{figure}

In Fig. \ref{Fig1}, variation of SI versus intracavity photon number and in terms of different settings for input parameters $\{\alpha,\beta,\vartheta\}$ is illustrated.

\subsection*{Operation Regimes}

Another very important result which can be drawn from the above discussions, is marking the boundaries of linear, weakly nonlinear, and strongly nonlinear interaction regimes in quantum optomechanics. This follows by normalizing $\delta$ with respect to the mechanical frequency $\Omega$ first, as $\bar{\delta}=\delta/\Omega$.  

\begin{itemize}[leftmargin=*]
	\setlength\itemsep{-0.4em}
	
	\item Fully Linear: This regime is easily given by $\bar{n}<<\bar{n}_{\rm max}$, where intracavity photon number is essentially too low to cause any appreciable SI. Here, the behavior of normalized SI is proportional to $\bar{n}$.
	
	\item Weakly Nonlinear: This regime is next given by $\bar{n}\sim\bar{n}_{\rm max}$ around the optimum operation point, where the SI rises to attain a maximum. The behavior of normalized SI is nearly Lorentzian centered at $\bar{n}=\bar{n}_{\rm max}$, with an intracavity photon number linewidth of $\Delta\bar{n}=\vartheta\bar{n}_{\rm max}$. 
	
	\item Stongly Nonlinear: This regime at larger intracavity photon numbers satisfying $\bar{n}>>\bar{n}_{\rm max}$ will push the system into strongly nonlinear regime where the SI quickly start to fade away. Here, the behavior of normalized SI is inversely proportional to $\bar{n}$.
\end{itemize}
These three behaviors in above operation regimes can be respectively displayed as
\begin{eqnarray}
\label{eq21}
\bar{\delta}(\bar{n})&\sim&\left(\frac{\bar{n}}{\bar{n}_{\rm max}}\right)\bar{\delta}_{\rm max}, \,\,\,\bar{n}<<\bar{n}_{\rm max},\\ \nonumber
\bar{\delta}(\bar{n})&\sim&\left\{1+\vartheta^{-2}\left(\frac{\bar{n}}{\bar{n}_{\rm max}}-1\right)^2\right\}^{-1}\bar{\delta}_{\rm max},  \,\,\, \bar{n}\sim\bar{n}_{\rm max},\\ \nonumber
\bar{\delta}(\bar{n})&\sim&\left(\frac{\bar{n}_{\rm max}}{\bar{n}}\right)\bar{\delta}_{\rm max}, \,\,\, \bar{n}>>\bar{n}_{\rm max}.
\end{eqnarray}

\subsection*{Fundamental Symmetries}

It is easy to verify that the SI does not violate the two fundamental symmetries of the nature. Here, both the time-reversal symmetry as well as the conservation of energy are preserved. The energy of scattered red- and blue- photons $\hbar\omega\mp\hbar\Omega$ is normally expected to be within the energy of one phonon $\hbar\Omega$ where $\omega$ is the angular frequency of incident electromagnetic radiation. Per every annihilated photon, exactly one phonon is either annihilated, giving rise to a blue-shifted photon, or one phonon is created, giving rise to a red-shifted photon.

However, not all phonons are having exactly the same energies. This is permissible by the non-vanishing mechanical linewidth $\Gamma>0$ of the cavity. One should expect that once this quantity vanishes, the SI is gone, since it is by (\ref{eq19}) proportional to $\Gamma^2$. Hence, basically it should be not contradictory to have a possible non-zero SI.

With regard to the time-reversal symmetry, one must take notice of the fact that all optical frequencies are physically positive, since we first must move back out of the rotating reference frame. For instance, the blue- and red-scattered photons have frequencies given by $\omega_b=\omega_c+\Omega-\frac{1}{2}\delta$ and $\omega_r=\omega_c-\Omega-\frac{1}{2}\delta$. Therefore, blue and red processes are not time-reversed processes of each other, as they both stay on the positive frequency axis. Negative frequency images corresponding to both processes do however exist and exactly satisfy the time-reversal.

\subsection*{Optomechanical Experiments}

To the best knowledge of the author, at least two very high resolution optomechanical experiments on side-band resolved samples, reveal the existence of SI, shown in Table \ref{Table1}. One experiment \cite{Schleisser} exhibits a normalized SI of $(3.5\pm0.9)\times 10^{-6}$, equivalent to $\delta=2\pi\times (142\pm36){\rm Hz}$ with $\Omega=2\pi\times40.5968{\rm MHz}$. There exists a much more recent experiment, where the observed normalized SI is much stronger $(9.23\pm0.66)\times 10^{-4}$ \cite{Vivishek}.

\begin{table}[ht!]
	\centering
	\begin{threeparttable}
		\caption{Side-band inequivalence in absolute values and normalized quantities in optomechanical experiments. Both experiments are carried out on side-band resolved cavities and have sufficiently high resolution to look for deviations from ideal expected case. Further examples are analyzed and discussed in Supplementary Information. \label{Table1}}
		\begin{tabular}{cccc}
			\hline\hline
			Experiment & SI $\delta/(2\pi)$ & Normalized SI $\bar{\delta}=\delta/\Omega$ & Remarks \\ \hline
			Schleisser \textit{et al.} \cite{Schleisser} & $142\pm36{\rm Hz}$ & $(3.5\pm0.9)\times 10^{-6}$ & Resonant pump\tnote{a} \\
			Sudhir \textit{et al.} \cite{Vivishek} & $3.97\pm0.284{\rm kHz}$ & $(9.23\pm0.66)\times 10^{-4}$ & Out-of-loop heterodyne spectra\tnote{b} \\
			\hline\hline
		\end{tabular}
		\begin{tablenotes}\footnotesize
			\item [a] Analysis done on Fig. 5b \cite{Schleisser} and the data corresponding to the side-bands at resonant $\Delta = 0$ pumping, which are much more visible and reliable. For near-blue resonant pump with $\Delta\approx-\Omega$ deviations lie within the error margin and are thus inconclusive. Coordinates of side-band Lorentzians were taken from ultra high-resolution digitalization of the graph. \item [b] Analysis done on the open-loop configuration with no feedback in Fig. 4b \cite{Vivishek}.
		\end{tablenotes}
	\end{threeparttable}
\end{table}

Inclusion of higher-order side-bands up to the $n-$th order would have caused an algebraic equation of the order $2n$ in place of the quadratic (\ref{eq12}), which is in general impossible to solve explicitly. Furthermore, the numerical advantage of treating such a generalized case would be only minor. With regard to the imaginary part $\Im[\delta]$ which was discarded, even if it were non-zero in the exact analysis, it could have resulted for instance in modification of linewidths. This calls for a deeper study of this subject, and it is also worth mentioning that the asymmetry of linewidths in optomechanics has recently been studied elsewhere in the linearized regime \cite{Niels}.

\subsection*{Raman Scattering}
Remarkably, the Stokes and anti-Stokes peaks of Raman spectra also exhibit the same phenomenon of SI. This has so far skipped the attention since normally the nonlinear interaction rate causing the formation of Raman scattering is too small for majority of bulk materials and liquids. Again, a full linear theory of Raman scattering \cite{Florian} disallows SI. However, with the recent advent of low-dimensional materials, optical nonlinear interactions leading to Raman scattering \cite{Fullerene,CNT,Springer,Raman2D,RamanGr} and Kerr effect \cite{CTP} are available at much stronger rates. 

Table \ref{Table2} provides a summary of measurable SI based on the reported Raman scattering of different materials. Calculations of normalized SI and errors are done according to the best possible resolution of measurement graphs. It has to be noted that for every given material, the SI is a strong function of scattering order, excitation wavelength, polarization, angle of incidence, as well as intensity. Therefore, it is not possible to reconstruct a fit to varying function for $\delta$ as (\ref{eq19}). However, the existence of a previously unnoticed SI, which is always leaning toward the red side-band seems to be conclusive. Therefore, a rigorous quantum theory of SI for Raman scattering from continuum optomechanics is yet to be developed, such as the one recently developed for materials \cite{Florian}. 

\begin{table}[ht!]
\centering
\begin{threeparttable}
\caption{Side-band inequivalence in Raman spectra of various materials, computed based on the highest-resolution available measurements. The last four rows under the single separating horizontal line are obtained by direct calculation on available high-resolution measurement data. Confidence interval for these last three measurements is remarkably significant, leaving no doubt in existence of SI. Further examples are analyzed and discussed in Supplementary Information. \label{Table2}}
	\begin{tabular}{ccc}
		\hline\hline
		Material & Normalized SI $\bar{\delta}=\delta/\Omega$  & Remarks \\ \hline
		$\text{MoTe}_{2}$ & $(7\pm 0.8)\times10^{-2}$ & 6-layer \cite{MoTe2} \\
		$\text{MoS}_{2}$ & $(3\pm 1)\times10^{-3}$ & 6-layer \cite{MoS2-1}  \\
		 & $(7.8\pm 1.5)\times10^{-3}$ & flake \cite{MoS2-2}  \\
		CNT\tnote{a} & $(6.0\pm 2.1)\times10^{-3}$ & $(10,5)$ SC\tnote{b} \, \cite{CNT1} \\
		 & $(2\pm 1)\times10^{-3}$ & 1st-order; SC\tnote{b} \, \cite{CNT2} \\
		 & $(1.22\pm 0.13)\times10^{-2}$ & 2nd-order; SC\tnote{b} \,\cite{CNT2} \\
		 & $(1.5\pm 0.18)\times10^{-2}$ & single-wall \,\cite{CNT3} \\
		 & $(2.11\pm 0.35)\times10^{-2}$ & single-wall \,\cite{CNT4} \\
		Gr\tnote{c} & $(8.0\pm 1.6)\times10^{-3}$ & 4-layer \cite{Graphene} \\
		 & $(5\pm 1.7)\times10^{-3}$ & 3-layer \cite{Graphene} \\
		 & $(1.63\pm 0.48)\times10^{-2}$ & $(1,2)$-twisted \cite{Graphene2} \\
		 & $(2.2\pm 1.09)\times10^{-2}$ & ${\rm C}_{31}$ $(1,3)$-twisted \tnote{d} \,\cite{Graphene2} \\
		 & $(3.6\pm 1.8)\times10^{-2}$ & ${\rm C}_{32}$ $(1,3)$-twisted \tnote{d} \,\cite{Graphene2} \\
		C & $(1.15\pm 0.57)\times10^{-2}$ & Bulk \cite{Graphene3} \\
		S\tnote{e} & $(1.03 \pm 0.25)\times10^{-2}$ & 1st-order; grains \cite{MoS2-2} \\
		 & $(7.2 \pm 1.8)\times10^{-3}$ & 2nd-order; grains \cite{MoS2-2} \\
		 Ethanol & $(1.8\pm 0.4)\times10^{-3}$ & 1st-order; EX\tnote{f} \, \cite{NanoPhoton} \\ \hline
		 ${\rm H}_2$ & $(4.7\pm 0.223)\times10^{-4}$ & MOM\tnote{g} \, \cite{MOM} \\ 
		 Gr & $(2.67\pm0.074)\times10^{-3}$ & $P_{\rm op}=7.0\text{mW}$\tnote{h} \,\cite{Tullio}\\
		 & $(1.67\pm0.165)\times10^{-3}$ & $P_{\rm op}=3.8\text{mW}$\tnote{h} \,\cite{Tullio}\\
         & $(0.67\pm0.603)\times10^{-3}$ & $P_{\rm op}=1.8\text{mW}$\tnote{h} \,\cite{Tullio}\\
		B4\tnote{i} & $(3.4$\tnote{j}\,\,$\pm0.18)\times10^{-3}$ & 1090nm line \cite{Pico,PicoData} \\
		SRA\tnote{k} & $(1.424\pm 0.01)\times10^{-3}$ & Yb-doped fiber  \cite{Fiber,FiberData} \\
		\hline\hline
	\end{tabular}
\begin{tablenotes}\footnotesize
	\item [a] CNT: Carbon Nano-tube.
	\item [b] SC: Semiconducting.
	\item [c] Gr: Graphene.
	\item [d] $1.96\text{eV}$ excitation.
	\item [e] S: Sulphur.
	\item [f] EX: Excitation at 532nm.
	\item [g] MOM: Molecular Optical Modulation; Measurement at $1015{\rm kPa}$ done on ortho$-{\rm H}_2$, which has a $17.6{\rm THz}$ rotational motion frequency. Experimental data received from the authors through private communication.
	\item [h] $3\text{ps}$ Ultrafast Excitation; Experimental data received from the authors through private communication. Data has to be shifted to adjust for $\delta(0)=0$, and exhibits a linear rate of $3.867\times10^{-4}/{\rm mW}$ versus input power, in agreement with the weakly nonlinear regime of (\ref{eq21}) as $\delta(\bar{n})\propto\bar{n}$.
	\item [i] B4: Single Biphenyl-4-thiol molecule in monolayer confined to optical picocavity; Experimental data available online \cite{PicoData}.
	\item [j] 4-point moving average applied. Guassian 10-point filtering results in the much larger value of $1.06\times10^{-2}$.
	\item [k] SRA: Stimulated Raman Amplifier; Experimental data available online \cite{FiberData}.
\end{tablenotes}
\end{threeparttable}
\end{table}

In the end and following the above, let us now plug-in (\ref{eq4}) and (\ref{eq7}) into the first of (\ref{eq10}). Some simplifications, while ignoring the inequilibrium quantum thermal effects on the population of side-bands, gives the equation
\begin{eqnarray}
\label{eq22}
\bar{n}_r-\bar{n}_b&\approx&\left(\frac{\Omega\bar{n}}{\Omega^2+\frac{1}{4}\Gamma^2}\right)\delta \approx\frac{\bar{n}}{\Omega}\delta.
\end{eqnarray}
\noindent
Here, $\bar{n}_r=|\braket{\bar{a}_r}|^2$ and $\bar{n}_b=|\braket{\bar{a}_b}|^2$ respectively refer to the number of scattered photons unto red and blue side-bands. Then, from (\ref{eq19}), and assuming that $\bar{N}=(\bar{n}_r-\bar{n}_b)/\bar{n}$ denotes the normalized asymmetry of side-bands, we get
\begin{equation}
\label{eq23}
\bar{N}(\bar{n})=\frac{\psi(1+\alpha\bar{n})}{\vartheta^2+(1-\beta\bar{n})^2}.
\end{equation}
Accordingly the asymmetry is increases up to a positive maximum, before decreasing back to zero at sufficiently high powers. 

\subsection*{Side-band Inequivalence in Other Areas}
Finally, it is easy to see that the same nonlinear symmetry breaking can lead to asymmetry in the particle pair production or parametric down conversion, which can be considered as the dual of optomechanical process \cite{Pair1,Pair2}. In order to observe this fact, consider an optomechanical system with a mechanical frequency roughly double the optical frequency $\Omega\approx2\omega$. If the mechanics is driven strong enough at the frequnecy $\Omega$, then the effective interaction Hamiltonian will be simply $\mathbb{H}_{\rm eff}=i\hbar g(\hat{a}^\dagger\hat{a}^\dagger\hat{b}-\hat{a}\hat{a}\hat{b}^\dagger)$, where a phonon with energy $\hbar\Omega$ dissociates into two photons with energies $\hbar(\omega\pm\delta)$ with $\delta$ representing the corresponding symmetry breaking in pair frequencies caused by SI. Parametric down conversion for phonons has recently been observed and reported, too \cite{Pair3}.

Also, based on the duality of effective interaction Hamiltonian in linear electro-optic modulation (within the validity of rotating wave approximation), with the optomechanical Hamiltonian \cite{EOV1,EOV2,EOV3}, one could predict that the same nonlinear inequivalence to appear in relevant experiments, too. As an example, optical modulation of Hydrogen \cite{MOM} is already shown in Table \ref{Table2} to exhibit this phenomenon. Hence, further implications could be expected in communications technology and filtering, where precise positioning of side-bands are of importance. Similar arguments should be valid for enhanced Raman scattering of single molecules by localized plasmonic resonances as well \cite{Roelli}.

\section*{Conclusions}

We presented a complete analysis of SI in quantum optomechanics, and showed it undergoes a maximum and obtained closed-form expressions for optimum intracavity photon number as well as maximum attainable SI. We classified the operation into the linear, weakly nonlinear, and strongly nonlinear regimes, in which the behavior of system is markedly different. The results of this investigation can provide the accuracy constraints as well as necessary experimental set up to resolve the elusive SI. Analysis of high resolution measurements of Raman scattering for different materials confirms the existence of SI. One could speculate that precise measurement of the variation of side-band inequialence in terms of various system parameters could provide further insight into unexplored nonlinear properties of different material.

\section*{Additional Information}

The author declares no competing interests.

\section*{Acknowledgment}

This paper is dedicated to the celebrated artist, Anastasia Huppmann.

\section*{Electronic Supplementary Material}

For an extremely comprehensive and categorized analysis of published experiments in quantum optomechanics, Brillouin scattering, Raman scattering, ion traps, Paul traps, and electrooptic modulation, the reader is referred to the Supplementary Information. Furthermore, in the nonlinear classical limit and based on analytical breathing solutions to optomechanics, existence of a positive shift in the spectrum can also be established mathematically, being consistent with the SI. These are presented in the Supplementary Information provided along with this article.

\setlength{\voffset}{0cm}
\setlength{\hoffset}{0cm}



\section*{Supplementary material}

# Supplementary Information: Further Evidence from Published Literature

## Analysis of Side-band Inequivalence
### Sina Khorasani

*"Plus un fait est extraordinaire, plus il a besoin d'être appuyé de fortes preuves; car, ceux qui l'attestent pouvant ou tromper ou avoir été trompés, ces deux causes sont d'autant plus probables que la réalité du fait l'est moins en elle-même."*
Pierre-Simon Laplace [S1]

## Abstract

This document includes a comprehensive and categorized referenced listing of published experiments which exhibit detectable Side-band Inequivalence (SI). Because of the resonant behavior of SI as found for $\delta$ in (14) of the main article, most experiments actually appear to have missed that resonance. Only those which hit the near-resonance criteria could have resulted in noticeable effect. Traces of SI could be found in a wide class of physically equivalent experimental platforms including Quantum Optomechanics, Ion and Paul Traps, Electrooptic Modulation, Brillouin Scattering, and Raman scattering. In what follows, we present an reasonably convincing overview of published available works, and divide them into various categories, and also discuss the existence of SI in approximate analytical solutions of classical optomechanics.

## Disclaimer

Contents of this document reflect the opinion of the author of present study (S.K.), and not necessarily those of original authors and/or laboratories, where respective works were carried out and here are cited. In some cases, the original authors were contacted and few agreed to the ultimate conclusions on SI drawn here while most remained unjudgemental. It is to be understood that referenced works contain public and extremely valuable information, and every effort has been made to ensure that the accuracy of analyses be kept as high as possible. Wherever raw measurement data has been available publicly or been graciously shared by the experimenters, calculations are carried out directly on numerical data. Otherwise, high-resolution digitization of graphs has been used to derive side-band inequivalences and frequencies to the best possible accuracy.

## S1 Quantum Optomechanics

The family of optomechanical interactions having the form $\mathbb{H}_i = \hbar g_0 \hat{a}^\dagger \hat{a} (\hat{b} + \hat{b}^\dagger)$ could be found in cavity quantum optomechanics, cavity quantum electromechanics, and quantum magnonics [S2–S5]. In all these experiments, interactions take place inside a doubly-confined cavity where both photons and phonons are kept interacting subject to decay. This platform provides the most straightforward and reliable source of SI as long as it is large enough to be measured. Typically, it is too small here and for all practical reasons, normalized SI in cavity quantum optomechanics is normally in the range of $10^{-6}$ to $10^{-4}$. This calls for very precise fabrication, calibration and stable measurements, which makes this type the most difficult way to obtain SI.

In cavity quantum optomechanics, there are three recognized and analyzed types of asymmetry between red and blue side-bands:

1. Side-band Asymmetry: which is a quantum effect connected to different thermal phonon occupation of side-bands [S2]. Under thermal equilibrium the red side-band is more populated since it has lower frequency.
2. Line-width Asymmetry: which is due to the different linewidths of red and blue side-bands. This phenomenon is also recently analyzed and now well understood [S7].
3. Side-band Inequivalence (SI): which is the topic of this article under consideration.

Optomechanical experiments may exhibit a tiny yet detectable SI [S8–S10], which is always biased towards red. A remarkable optomechanical experiment [S9] on an optomechanical crystal reports precise measurement of side-bands at various pump levels without and with cooling tone. In absence of cooling tone, the location of side-bands asymmetrically moves almost linearly with the measured photon population, up to 4kHz [S9]. Another very high precision experiment [S8] on micro-toroids, exhibits markedly frequency asymmetry both in transmission and power spectral density of side-bands. The difference in particular becomes quite noticeable for the second-order side-bands, which do not quite fit well to the linearized model. The SI here is around 142Hz.

There exist another report in solid-state cavity quantum optomechanics with sufficiently high resolution [S10] and right conditions which exhibits SI, too. Under thermal equilibrium and absence of cooling tone, the side-bands are well resolved and sharp enough to detect frequency differences, and again here SI is biased towards red. Analysis of a report on motional side-bands in solid-state optomechanical cooling by controlling the interaction with surface acoustic waves [S11] in a similar manner shows a noticeable SI $\bar{\delta} \approx 0.74\%$, which is of course biased towards red.

It should be noted that optically confined particles in side-band resolved regime can display remarkably strong SI. Given the fact that the nature of these experiments are markedly different than cavity quantum optomechanics, it is quite exciting indeed to observe the fact that SI does survive. A recent such experiment [S12], with the reported uncertainties taken into account, exhibits an SI as large as $\bar{\delta} = 0.8\%$, if we assume that the zero calibration is sufficiently more accurate than 0.1 kHz. Otherwise, referencing with respect to the fitted central resonance gives $\bar{\delta} = 0.43\% \pm 0.18\%$. In any case, it is positive as expected within a reasonable confidence interval, that is the red peak is definitely further away from the resonance as opposed to the blue. For this experiment, the raw measurement data was made available to the author, providing an accurate analaysis, as depicted in Fig. S1. Optical confinement of a single-atom using tweezers [S13] also leads to measurement of single-atom sideband spectra,

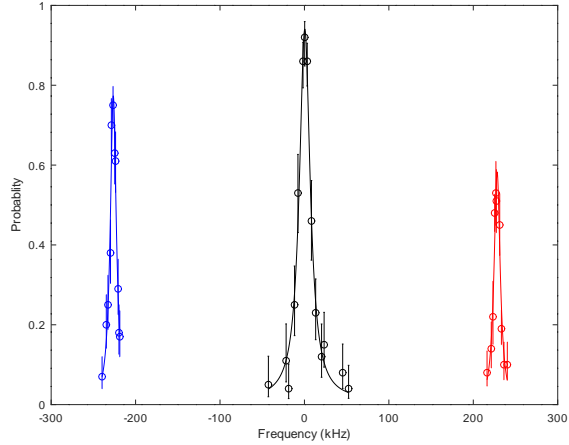

**Figure S1.** Motional sidebands of an optically trapped ion in a tweezer before sideband cooling obtained from spectroscopic raw measurement data [S12].

which in absence of cooling can be measured to have a normalized SI of $\bar{\delta} = 1.04\%$.

In a very recent landmark experiment carried out [S14] side-bands were generated as a result of controlled detuned acoustooptic modulation, where detuning frequency effectively replaces modulation frequency. The authors have made the measurement data available online [S15] from which the SI may be calculated with respect to the modulation frequency. The behavior of SI here is dependent on the parameters of modulator and not the optomechanical device, so we had to look for fitting parameters. The appropriate expression in the limit of very small $\Gamma$ follows (17) in the main article. Assuming $\gamma \approx \kappa$ and $g = g_0\sqrt{\bar{n}}$, it takes the form

$$\bar{\delta}(\Omega) = \frac{8g^2}{\kappa^2} \frac{\Omega^2 \mathrm{sgn}[\Omega]}{\Omega^2 + 4\frac{1}{\kappa^2}\left(\Omega^2 - 2g^2\right)^2}, \tag{S1}$$

noticing that $\Omega$ represents the detuning frequency and can be positive or negative. Hence, the normalized SI $\bar{\delta}$ may be considered as a function of $\Omega$, too, and $\mathrm{sgn}(\cdot)$ is the sign function. This expression (S1) very well fits the calculated behavior based on the analysis of raw measurement data [S15], giving a remarkably nice fit. However, it has to be pointed out again that the optimum fitting parameters $(\kappa, g) = 2\pi(620, 90)$kHz are irrelevant to the optomechanical system parameters.

### Breathing Solutions

In the fully classical approximation, the nonlinear optomechanical equations read

$$\frac{d}{dt}\Upsilon = \left(i\Delta - \frac{1}{2}\kappa\right)\Upsilon + ig_0\Upsilon(\Phi + \Phi^*), \tag{S2}$$

$$\frac{d}{dt}\Phi = \left(-i\Omega - \frac{1}{2}\Gamma\right)\Phi + ig_0\Upsilon^*\Upsilon.$$

Under the assumption of slowly varying amplitude $|d\Upsilon/dt| << \Omega|\Upsilon|$ and large optical quality factor $\kappa << |\Delta|$, the set of equations (S2) admits an analytical solution for the optical field of the form

$$\Upsilon(t) = \frac{\theta e^{i\theta^2\tau}}{\sqrt{2D/|\Delta|}}\left[\frac{2t^2\cosh(\vartheta\tau) - i2t\vartheta\sinh(\vartheta\tau)}{\cosh(\vartheta\tau) - \sqrt{1 - t^2}} - 1\right], \tag{S3}$$

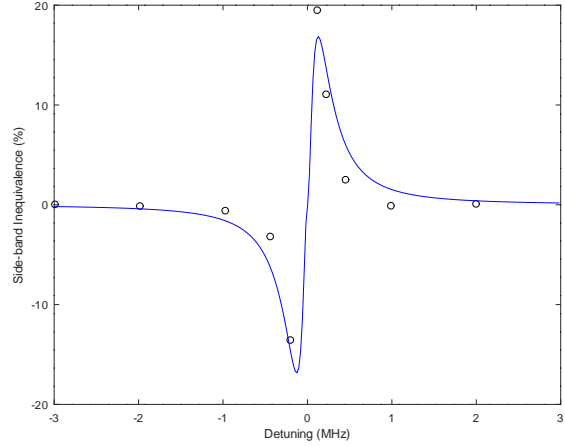

**Figure S2.** Calculated SI from raw measurement data [S15] and fitting function based on (S1) with modulator parameters $g = 2\pi \times 90 \text{kHz}$ and $\kappa = 2\pi \times 620 \text{kHz}$.

referred to Akhmediev breathers [S16, S17]. Here, $\tau = |\Delta|t$ is normalized time, $\theta$ is a real constant to be determined, $D$ is defined in (15), and $\vartheta = -2\theta^2 \iota \sqrt{1 - \iota^2}$. For $\iota < 1$ solutions are breathing soliton-like, while for $\iota > 1$ solutions become nonlinear oscillatory waves.

When $\iota < 1$, then the expression within the brackets is composed of an even real part and an odd imaginary part. Hence, the Fourier transform of this hyperbolic expression within the brackets is real-valued, which we here denoted by $\mathscr{G}(w)$. Obviously, the magnitude of Fourier transform, or the spectrum $|\mathscr{G}(w)|$ no longer needs to be even. Hence, the Fourier transform of $\Upsilon(t)$, which is given by $\mathscr{I}(w) = \mathscr{F}\{\Upsilon(t)\}(w)$ is simply the shifted transform $\mathscr{G}(w - \theta^2|\Delta|)$, and the amount of frequency shift is nothing but the SI. This fact is a direct result of the multiplying term $\exp(i\theta^2\tau)$ in the analytical solution (S3), and thus the SI.

While, the algebraic hyperbolic form of (S3) with $\iota < 1$ disallows analytical evaluation of the spectrum $\mathscr{I}(w)$, however, calculation of $\mathscr{I}(w)$ from (S3) with $\iota > 1$, which leads to periodic nonlinear oscillatory solutions, becomes explicitly possible as shown in the following. To do so, we may proceed with the substitution $\iota = |(\Omega + i\Gamma/2)/g| \approx \Omega/g$ where $g = g_0\sqrt{n}$ is the enhanced optomechanical interaction rate, which in the weakly-coupled limit obviously satisfies $\iota \gg 1$. Hence, we have $i\vartheta = 2\theta^2\iota\sqrt{\iota^2 - 1} \approx 2\theta^2\iota^2$. Now, we can rearrange (S3) as

$$\Upsilon(t) = \frac{2\iota\theta e^{i\theta^2\tau}}{\sqrt{2D|\Delta|}} \left[ \frac{\iota|\Delta|\cos(\varpi t) + i\varpi \sin(\varpi t)}{\cos(\varpi t) - i\sqrt{\iota^2 - 1}} - \frac{|\Delta|}{2\iota} \right],$$ (S4)

where the replacement $i\vartheta\tau = \varpi t$ has been made and $\varpi = \Omega + \delta\Omega$ is the shifted mechanical frequency while taking the optical spring effect $\delta\Omega$ into account. The expression within brackets in (S4) can be simplified using trigonometric identities as

$$\begin{aligned}
\Upsilon(t) &= \Xi e^{i\frac{1}{2}\delta t} \left[ \frac{(\iota|\Delta| + \varpi)e^{i\varpi t} + (\iota|\Delta| - \varpi)e^{-i\varpi t}}{e^{i\varpi t} + e^{-i\varpi t} - 2i\sqrt{\iota^2 - 1}} - \frac{|\Delta|}{2\iota} \right] \\
&= \Xi e^{i\frac{1}{2}\delta t} \left[ \frac{2i\sqrt{\iota^2 - 1}(\iota|\Delta| + \varpi) - 2\varpi e^{-i\varpi t}}{e^{i\varpi t} + e^{-i\varpi t} - 2i\sqrt{\iota^2 - 1}} + \frac{2\iota(\iota|\Delta| + \varpi) - |\Delta|}{2\iota} \right] \\
&= 2\Xi e^{i\frac{1}{2}\delta t} \left[ \frac{i\varsigma - \varpi e^{-i\varpi t}}{e^{i\varpi t} + e^{-i\varpi t} - 2i\sqrt{\iota^2 - 1}} \right] + \Xi \left[ \frac{2\iota(\iota|\Delta| + \varpi) - |\Delta|}{2\iota} \right] e^{i\frac{1}{2}\delta t} \\
&= i\Psi e^{\frac{1}{2}\delta t} f(t) + \Lambda e^{i\frac{1}{2}\delta t},
\end{aligned}$$ (S5)

in which $\Xi = \sqrt{2}\iota\theta/\sqrt{D}$, $\Lambda = \Xi(2|\Delta|\iota^2 - 2\varpi\iota - |\Delta|)/2\iota$, $\varsigma = \sqrt{\iota^2 - 1}(\iota|\Delta| + \varpi)$, and $\Psi = 2\Xi\varsigma$. We here may notice that $\delta = \theta^2|\Delta| > 0$ is nothing but the SI. The function $f(t)$ can be rewritten as

$$\begin{aligned}
f(t) &= \frac{1 + i\frac{\varpi}{\varsigma}e^{-i\varpi t}}{e^{i\varpi t} + e^{-i\varpi t} - 2i\sqrt{\iota^2 - 1}} = \frac{e^{i\varpi t} + i\varkappa}{(e^{i\varpi t} - i\sqrt{\iota^2 - 1})^2 + \iota^2} \\
&= \frac{e^{i\varpi t} + i\varkappa}{(e^{i\varpi t} - i\zeta^+)(e^{i\varpi t} - i\zeta^-)} = \frac{1}{2\iota}\frac{\zeta^+ + \varkappa}{e^{i\varpi t} - i\zeta^+} - \frac{1}{2\iota}\frac{\zeta^- + \varkappa}{e^{i\varpi t} - i\zeta^-} \\
&= h(e^{i\varpi t}),
\end{aligned}$$ (S6)

where the replacement $\zeta^\pm = \pm \imath + \sqrt{\imath^2 - 1}$ has been made and $\varkappa = \varpi/\varsigma$ with $\zeta^+ \zeta^- = -1$, $\zeta^+ > 0$ and $\zeta^- < 0$ respectively lying in the upper and lower half-complex planes. Obviously, the pole at $\imath \zeta^+$ falls outside the unit circle, and only the pole at $\imath \zeta^-$ remains inside the unit circle. Hence, $\varkappa << -\zeta^- << 1 << \zeta^+$. These will be needed later to derive the spectrum of (S4).

All remains now is to solve $\imath \vartheta \tau = \varpi \imath$, which gives rise to the approximate solution

$$\delta \approx \frac{2\varpi}{\imath \sqrt{\imath^2 - 1}} = 2g^2 \frac{\Omega + \delta\Omega}{\sqrt{\Omega^2 + \frac{1}{4}\Gamma^2}\sqrt{\Omega^2 + \frac{1}{4}\Gamma^2 - g^2}}. \tag{S7}$$

This can be further written as

$$\frac{\delta}{2} \approx \frac{\Omega + \delta\Omega}{\imath^2} = g^2 \frac{\Omega + \delta\Omega}{\Omega^2 + \frac{1}{4}\Gamma^2} \approx \frac{g^2}{\Omega} + \frac{g^2}{\Omega^2}\delta\Omega \approx \frac{g^2}{\Omega} - \frac{2g^4\Delta}{\Omega^2}\frac{(\Omega^2 - \frac{1}{4}\kappa^2)}{(\Omega^2 + \frac{1}{4}\kappa^2)^2}, \tag{S8}$$

since $\delta\Omega \approx -2g^2\Delta(\Omega^2 - \frac{1}{4}\kappa^2)/(\Omega^2 + \frac{1}{4}\kappa^2)^2$. It is here seen that the correction arising from spring effect is $O(g^4)$ while the SI $\delta$ is $O(g^2)$. Hence, the normalized SI $\bar{\delta} = \delta/2\Omega$ can be finally approximated for the side-band resolved regime $\Omega >> \kappa$ as

$$\bar{\delta} \approx \frac{2g^2}{\Omega^2}\left[1 - \frac{2g^2\Delta(\Omega^2 - \frac{1}{4}\kappa^2)}{\Omega(\Omega^2 + \frac{1}{4}\kappa^2)^2}\right] \approx \frac{2g^2}{\Omega^2}\left[1 - \frac{2g^2}{\Omega^3}\Delta\right], \tag{S9}$$

showing that the correction of optomechanical spring effect to the ansatz (1), ultimately yielding the expression for SI (14) had been safely ignored indeed. There is also a higher-order correction to the optomechanical spring effect $\delta\Omega$ [S2] as a result of non-zero coherent phonon population $\bar{m}$, which results in an extra correction to (S8) by replacing $g = g_0\sqrt{\bar{n}}$ within the brackets of (S9) with $g \approx g_0\sqrt{\bar{n} + \bar{m} + 1}$. But this leaves our derivations and conclusions regarding the SI unchanged.

This solution (S5) is a bi-periodic and complex-valued product of two periodic functions $f(t)$ and $\exp(i\delta t)$. Fourier transform of the periodic function $f(t) = f(t + 2\pi/\varpi)$ defined as $F(w) = \frac{1}{2\pi}\int_{-\infty}^{\infty} f(t)e^{-iwt}dt$ is straightforward to obtain. In fact, one should have $F(w) = \sum_{\nu} f_\nu \delta(w - \nu\varpi)$, where $\delta(\cdot)$ are Dirac's delta functions, and $f_0 = f(1)$ while $f_\nu = \exp(i2\pi/\nu)$, $\nu \neq 0$. Hence, it is straightforward to see that how its spectrum looks like. Defining the spectrum of $f(t)$ as $|F(w)|$ in the Fourier domain $w$ simply is $F(w) = \sum_\nu |f_\nu|\delta(w - \nu\varpi)$, which consists of Dirac deltas at $w = \pm\nu\varpi \approx \pm\nu\Omega$, $\nu \in \mathbb{N}$ corresponding to the side-bands. A practical system obviously does not exactly follow the breather solution (S3) and hence side-bands all have finite non-zero linewidths. Ultimately, we have $I(w) = |\mathscr{I}(w)|$.

Therefore the ultimate spectrum of the cavity within the approximation of breather solutions is $I(w) \approx |F(w - \frac{1}{2}\delta)| + R(w)$, where $R(w)$ is the reflection from cavity at central resonance $w = 0$. Since the reflected central resonance $R(w)$ normally masks out the zeroth harmonic, therefore the side-bands appear to be positioned asymmetrically in frequency equal to the SI $\delta$. Hence, $I(w)$ may be written conveniently as

$$\begin{aligned}
I(w) &= |\tfrac{1}{2\pi}\Lambda + i\Psi f_0|\delta(w - \tfrac{1}{2}\delta) + \Psi \sum_{\nu=1}^{\infty} \left[|f_\nu|\delta(w - \tfrac{1}{2}\delta - \nu\varpi) + |f_{-\nu}|\delta(w - \tfrac{1}{2}\delta + \nu\varpi)\right], \tag{S10} \\
f_\nu &= \frac{1}{2i\varpi\pi}\oint \frac{h(z)}{z^{\nu+1}}dz, \\
&= -\left[\frac{\zeta^- + \varkappa}{2\varpi\imath(i\zeta^-)^{\nu+1}}\right] + \frac{1}{\varpi\nu!}\left[\frac{d^\nu}{dz^\nu}h(z)\right]_{z=0}u(\nu) \\
&= \left[\frac{i(\zeta^- + \varkappa)}{2\varpi\imath\imath^\nu(\zeta^-)^{\nu+1}}\right] - \frac{\zeta^- + \varkappa}{2\imath\varpi\nu!}\left[\frac{(-1)^\nu\nu!}{\imath^{\nu+1}(\zeta^-)^{\nu+1}}\right]_{z=0}u(\nu) \\
&= \left[\frac{i(\zeta^- + \varkappa)}{2\varpi\imath\imath^\nu(\zeta^-)^{\nu+1}}\right][1 + (-1)^\nu u(\nu)].
\end{aligned}$$

where the change of variables $z = \exp(i\varpi t)$ has taken place, and the integration is taken counter-clockwise on the unit circle in the complex $u$-plane. The only contributing pole of $h(z)$ is at $z = i\zeta^- \approx -i/2\imath^2$. Furthermore, $u(\cdot)$ is the unit-step function, which allows the second term to contribute only if $\nu \geq 0$. The function $h(z) = (z - \varkappa)/(z - i\zeta^+)(z - i\zeta^-)$ was also defined in (S6). In (S10), the positive odd harmonics identically vanish, and $f_\nu$ and $f_{-\nu}$ respectively correspond to Stokes and anti-Stokes amplitudes. There are no odd-ordered Stokes components in the breather nonlinear oscillatory wave (S4), and also anti-Stokes components diminish in strength with their order $-\nu$ increasing as $(\zeta^-)^{-\nu}$ according to (S10).

The total power $P = \int_{-\infty}^{\infty} I(w)dw$ is now simply $P \approx \sum_\nu |f_\nu|$, and total harmonic distortion shall be given by the simple expression THD $\approx \sum_{|\nu|\geq 2} |f_\nu|/\sum_{|\nu|\geq 1} |f_\nu|$. The first-order mechanical side-bands correspond to $f_{\pm 1}$ with sharp peaks located at $\pm\varpi - \frac{1}{2}\delta \approx \pm\Omega - \frac{1}{2}\delta$, confirming the initial ansatz (1) and speculation regarding the existence of SI. In summary, the breathing

analytical solution ([S3](#)) actually highlights the existence of a non-zero SI, simply because of the multiplying term $\exp(i\theta^2\tau)$ and its bi-periodic form, and furthermore SI has to be always towards red (Stokes) simply because $\theta^2 > 0$ is always positive.

The ratios of coefficients $f_\nu$ in ([S10](#)) can be estimated using binomial expansion of denominator in ([S5](#)) and the original form, resulting in approximate expressions for the ratios of side-band powers. For instance, the ratio of optical amplitude in the first-order side-bands with respect to the central resonance is roughly

$$
\begin{aligned}
\frac{I_1}{I_0} &= \frac{|f_1| + |f_{-1}|}{2\left|\frac{1}{2\pi}\Lambda + i\Psi f_0\right|} = \frac{\pi}{\varpi\iota}\frac{\zeta^- + \varkappa}{\left[\Lambda - (\zeta^- + \varkappa)\pi\Psi/(\varpi\iota\zeta^-)\right]} \\
&\approx \frac{\pi\zeta^-}{\varpi\iota\Lambda - \pi\Psi}.
\end{aligned}
\tag{S11}
$$

The accuracy of breathing solutions for second- and higher-order harmonics is insufficient to obtain a meaningful ratio such as ([S11](#)), nevertheless, it exhibits a positive and unmistakable SI towards red.

## S2  Ion/Paul Traps

The volume of existing literature on atomic and ion traps is truly vast, and we limit the study to a collection of selected works in this area. SI numbers are typically large and quite noticeable.

Detection of motional side-bands around 5GHz in a linear Paul trap placed on an optical cavity [S18] gives rise to $\bar{\delta} = 1.32\% \pm 0.32\%$ at anti-node and $\bar{\delta} = 1.42\% \pm 0.24\%$ at node. Another measurement [S19] clearly shows red and blue side-bands separately from which one may obtain the fairly accurate estimation of $\bar{\delta} = 0.021\% \pm 0.0021\%$. Doppler cooling on a microchip multi-segmented ion trap [S20] gives $\bar{\delta} = 0.047\% \pm 0.0078\%$. Doublet features of side-bands are very much visible in the next research [S21], which was the primary assumption as displayed in (1) at the beginning of our analysis. One report considers a high-resolution measurement of motional side-bands of trapped $Ca^+$ ions [S22], which contains four very sharp resonances on either side. All of these resonances exhibit significant SI and the values are $\bar{\delta} = \{1.8\% \pm 0.29\%, 1.3\% \pm 0.21\%, 0.50\% \pm 0.12\%, 0.87 \pm 0.11\%\%\}$ sorted in terms of increasing shift frequencies, shown in Fig. [S3](#). Planar and vertical modes in ion traps can also individually have SI, as the recent measurements [S23] may give the respective values $\bar{\delta} = 0.096\%, 0.18\%$ within $\pm 0.008\%$. Axial motional sidebands in another microfabricated ion trap design [S24] are measured from which $\bar{\delta} = 0.037\% \pm 0.0085\%$ can be computed.

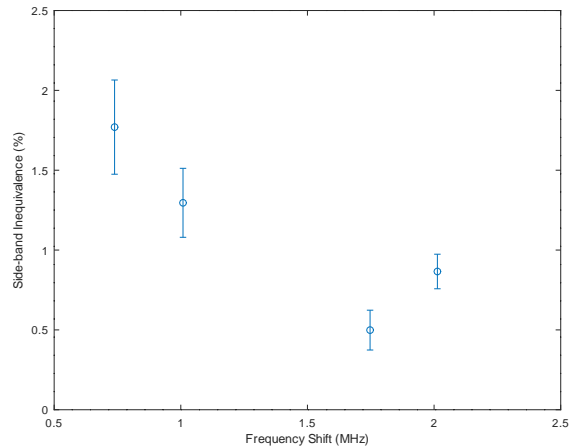

**Figure S3.** Side-band Inequivalences computed from digitization of microfabricated ion trap experiment [S22].

A very large SI also can be noticed in the side-bands of a calcium ion confined in a linear Paul trap [S25] as $\bar{\delta} = 6.3\% \pm 0.18\%$. Doppler cooling does not seem to remove this effect, as the value $\bar{\delta} = 0.10\% \pm 0.0070\%$ corresponds to cooled vibrational states of trapped atomic ion [S26]. Motional side-bands of two $_{40}Ca^+$ ions before sympathetic ground-state cooling [S27] show a clear SI as large as $\bar{\delta} = 0.35\% \pm 0.0030\%$. Also, an article [S28] which reports cooling of trapped $_{111}Cd^+$ ions shows $\bar{\delta} = 0.21\% \pm 0.0086\%$.

Finally, a very recent experimental study at Max-Planck Institute [S29] reports resolved-sideband cooling of an optical lattice with zoomed-out side-bands both before and after cooling takes place. The SI is so large that can be easily seen on the frequency scales equal to $\bar{\delta} = 1.9\%$. It turns out that the authors had noticed this large difference and had tried to explain it using anharmonicity of the confining potential [S30] in weak limit. It has to be mentioned that if the cause of side-band

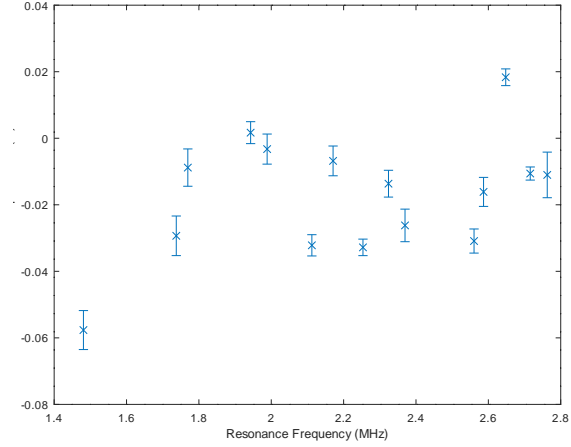

**Figure S4.** Side-band Inequivalences of an ion trap measured for the radial mode spectrum after Doppler cooling, computed from raw measurement data [S31].

inequivalence were anharmonicity, then the sign of expression for SI must follow the sign of anharmonic potential, that is $\delta \propto g_0$. However, the analysis in the current study shows that $\delta \propto g_0^2$ holds to a good accuracy, regardless of the sign of anharmonic optomechanical interaction.

Another article, the measured radial mode spectrum of a nine-ion linear crystal has been reported [S31] and the raw measurement data were made available to the author. There are multiple resonances on either of the blue and red sides, which are measured both after Doppler cooling and after Doppler/EIT cooling. After EIT cooling, the red side-bands almost disappear, and it has been possible to find the SI only after Doppler cooling, shown here in Fig. S4. The behavior of SI, while clearly non-zero, does not conform to the expectation and that could possibly because of Doppler cooling which dislocates resonances a bit. Nevertheless, nearly all of the 15 resonances contain non-zero SI.

Shown in Fig. S5, the SI of the motional state of a trapped $_{171}\mathrm{Yb}^+$ ion placed in a state-dependent potential generated by a running optical lattice is computed from raw measurement data, made available to the author [S32]. Measurements here correspond to the absence of cooling of any type. Every point is furthermore the average of 100 traces, and that provides an accurate estimation of SI. The maximum normalized SI is about $\bar{\delta} = 0.99$.

Interestingly, in the limit of small optomechanical interaction rate $g_0$ and low optical power, one may use (18) to obtain the approximate proportionality dependence

$$\bar{\delta}(\Omega) \propto \frac{1}{\Omega^2 + \frac{1}{4}\gamma^2}, \tag{S12}$$

where $\Omega$ and $\gamma$ are respectively the mechanical/modulation frequency and optomechanical decay rate. This information can be used now to check whether SI decreases with frequency in accordance to (S12). This is shown in red in Fig. S5.

As the last remark of this section, not all reported measurements apparently show SI in the desired way [S33, S34]. What these works have in common is that they include measured side-bands before (and after) cooling, and while the measureable SI seems significant, however, it is in the opposite direction. Normally, it is convenient to see the central resonance to make sure zero-referencing is accurately done, and it has come to the attention of the author that in most cases and in particular for Raman and Brillouin scattering measurements, the accurate SI cannot be found without such referencing. Anyhow, these three articles remain as unanswered questions, which need further investigation.

## S3 Electrooptic Modulation

One of the pioneering works on optical fibers [S35, S36] reports the optical intensity spectra scanned with a Fabry–Perot analyzer of a single frequency laser modulated by an electrooptic modulator. It is straightforward to identify an SI as large as $0.85\% \pm 0.2\%$, due to a modulation frequency $f_m$ on a carrier with frequency $\nu_0$ therein.

Similarly, one may find an SI as large as $2.20\% \pm 0.27\%$ and $0.42\% \pm 0.14\%$ for the depolarized spectrum of an LiNbO$_3$ cell [S37] held at $45°$ and modulated at respectively at $f_1 = 3\mathrm{GHz}$ and $f_1 = 8\mathrm{GHz}$.

It can be easily verified here that whether (S12) holds, as the normalized SI should roughly decrease with the second power of modulation frequency (or mechanical frequency where relevant). This actually happens to be the case since one

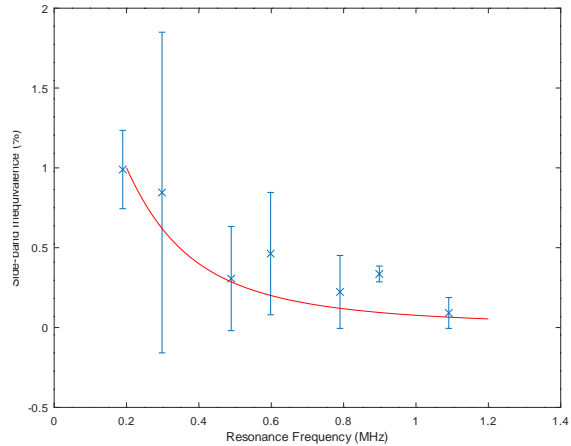

**Figure S5.** Side-band Inequivalences of an ion trap measured for the radial mode spectrum after Doppler cooling, computed from raw measurement data [S32]. The red fit is according to (S12) with $\gamma = 2\pi \times 63.6\text{kHz}$.

would expect then that $\bar{\delta}(f_1)/\bar{\delta}(f_2) \approx (f_2/f_1)^2$ hold. Checking the either side while taking care of error bounds gives $\bar{\delta}(f_1)/\bar{\delta}(f_2) = 0.2 \pm 0.09$ which contains $(f_2/f_1)^2 = 0.14$. These numbers are obtained from graph digitization, while such an agreement is later observed to a much higher accuracy for the case of Raman scattering where direct raw measurement data has been available.

The next reference [S38] is a thesis reporting the spectra of the Stokes/anti-Stokes signals at the output of a single-mode fiber which has been 50km long, modulated at 10.94GHz using an optical QPSK (Quadrature Phase-Shift Keying) modulator. The first three Stokes and anti-Stokes resonances are sharply measured and clearly visible, which is consistent with an unnormalized and positive SI of $\delta = (250 \pm 25)\text{MHz}$.

A recent article [S39] reports measurements of the optical spectrum of a phase modulated signal and the measurement data were made accessible to the author. The modulation provided up to seven side-bands on either side, which made the evaluation of SI possible. One would expect, likewise, that the largest SI would go to the lowest order, and that happens to be the case where an SI as large as 1.94% has been observed, which is shown in Fig. S6. There is yet another high resolution experiment on electrooptic modulation of light at telecommunications wavelength at 1545.91nm using a LiNbO$_3$ modulator with various modulation frequencies, where raw measurement data were made available [S40]. Shown in Fig. S7, the largest SI happens to occur for the lowest modulation frequency, and is very large up to 9.4%, indeed. Pressurized hydrogen as a result of molecular optical modulation [S41] has been shown to generate sidebands, due to ultrafast variation of molecule polarizability arising from coherent molecular motion. The author was also given access to the raw measurement data which was noticed to give rise to an SI of $\bar{\delta} = 0.047\%$.

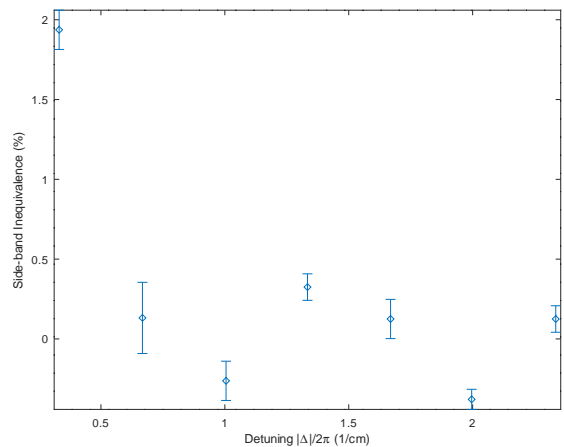

**Figure S6.** Side-band Inequivalence of an electrooptically modulated light computed from raw measurement data [S39].

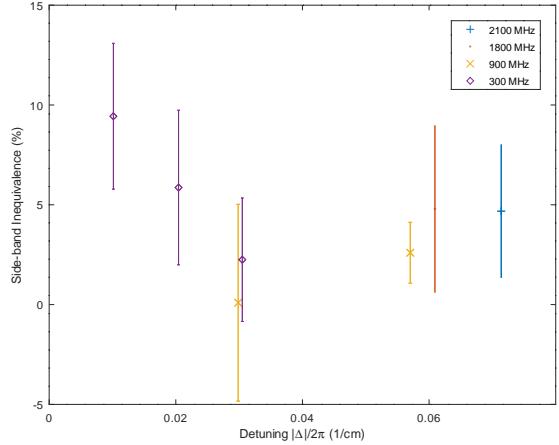

**Figure S7.** Side-band Inequivalence of an electrooptically modulated light computed from raw measurement data [S40].

It has to be stressed out that not every source of nonlinearity (such as cross-Kerr) or multiplying element could exhibit SI. It necessarily must be an Electrooptic medium such as externally driven LiNbO$_3$ cell or equivalent, with interaction Hamiltonian of the desired type. It remains as a subject for next studies to investigate the excessive distortion caused by this type of modulation and its implications on microwave and optical communication technologies.

## S4 Brillouin Scattering

In the contexts of Brillouin Scattering (BS) and Stimulated Brillouin Scattering (SBS), SI appears very clearly which is typically easy to measure due to sharp resonances, and it could take on large magnitudes up to a few percent. One study [S42] has clearly zoomed out the Stokes and anti-Stokes resonances at various power levels, which enables to examine the variation of SI with respect to optical power to a great precision. The measured SI for illuminating optical powers of $P_{op} = \{12, 14, 20\}$mW is respectively $\bar{\delta} = \{0.059\%, 0.059\%, 0.035\%\}$ showing that an optimal operation point around $(P_{max}, \bar{\delta}_{max}) = (13\text{mW}, 0.059\%)$ has indeed been hit. Had the authors taken more measurement points above and below this power level, it would have been possible to reconstruct the accurate dependence. These SI numbers are accurate better within than an error margin of 0.017%.

The next article [S43] reports a very high-resolution spectrum of pump together with the blue and red side-bands in an optical fiber due to Brillouin scattering. The corresponding wavelengths are also clearly marked, enabling one to easily calculate the SI after conversion of wavelengths to frequencies. Once done, the resulting SI is $\bar{\delta} = 0.027\%$ which leans towards the red side-band as it should be. Similarly in another report [S44], the SI can be found in spectra of the transmitted light through a single-mode optical fiber $3.0\% \pm 0.21\%$, which should be evaluated with respect to the dislocated central peak.

Brillouin scattering at various angles relative to a SiOC:H/Si film has been measured and reported [S45], which shows peaks for longitudinal guided and bulk modes. Here, the SI for longitudinal bulk mode is $\bar{\delta}(\theta = 65°) = 0.69\%$ and $\bar{\delta}(\theta = 25°) = 0.93\%$ where $\theta$ is the illumination angle, which more or less preserves its frequency regardless of $\theta$. The situation for guided modes is a lot more complex since it appears to be combined with the dispersion of modes like Raman scattering from Dirac materials (to be discussed next sub-section). Anyhow, one may observe approximately that $\bar{\delta}(\theta = 65°) = -1.43\%$ and $\bar{\delta}(\theta = 25°) = -5.01\%$ hold for longitudinal guided modes.

Brillouin scattering from surface acoustic modes of ZnO thin films grown on Si substrate is discussed in the next article [S46]. Two sets of measurements are done versus various angles of incidence $\theta = \{30°, 45°, 50°, 60°\}$ at fixed thickness of $h = 100$nm as well as different film thicknesses $h = \{20, 44, 100, 200, 320\}$nm at fixed incidence angle of $\theta = 45°$. All measurements lead to definite SI considering the errorbars. In these traces, the fundamental Rayleigh wave appears in common for which it is possible to get $\bar{\delta} = \{0.99\%, 0.73\%, 3.8\%, 2.5\%\}$ as well as $\bar{\delta} = \{3.16\%, 0, 2.5\%, 6.8\%, 5.3\%\}$ within the rough error margin of $\pm 0.5\%$.

Backward Brillouin scattering may also give rise to SI. This fact may be verified on the measurements carried out in an ultra-high $Q$ resonator [S47] which exceptionally the Stokes/anti-Stokes pair could be seen visibly. For this case, a value of $\bar{\delta} = -2.4\% \pm 0.6\%$ can be estimated, where the negative sign results from backward configuration which is equivalent to the replacement $\Omega$ by $-\Omega$.

A more recent research [S48] concerns Brillouin forward and backward scattering in various materials under two $(s, p)$ orthogonal polarizations. For the case of microscope slab in forward scattering two resonances $(L, T)$ can be distinguished

for each of $(s, p)$ orthogonal polarizations, and the resulting numbers for SI mark significant deviations from zero. For $p-$polarization, we have $\bar{\delta} = \{3.2\%, 2.0\%\}$ corresponding to the two resonances while for $s-$polarization, we have $\bar{\delta} = \{4.2\%, 0.83\%\}$ all within $\pm0.14\%$. It is instructive to examine the measurements corresponding to the forward and backward configurations which are both available for the case of cover glass. Both of the two resonances $(L, T)$ are sharp enough in the forward configuration for both $(s, p)$ polarizations, while in the backward configuration only one resonances can be clearly seen. The distinguishable resonance/polarization pairs here are $(L; s)$ and $(T; p)$. In the forward configuration we have $\bar{\delta}(L; p) = 0.28\%$ $\bar{\delta}(T; s) = 1.87\%$ both within $\pm0.14\%$. Meanwhile, the SI for forward modes are $\bar{\delta}(T; p) = 1.87\%$, $\bar{\delta}(L; s) = 0.83\%$ within $\pm0.14\%$ and for backward modes are $\bar{\delta}(T; p) = -0.37\%$, $\bar{\delta}(L; s) = -0.67\%$ within $\pm0.09\%$. Hence, it can be confirmed well that the SI for the backward Brillouin scattering can assume negative values, as we have verified for two very different types of measurements [S47, S48].

The comprehensive set of raw measurement data of hypersonic Brillouin scattering of surface acoustic waves in bulk transparent materials has been also made available to the author [S48]. These contain a total of 12 traces shown in Fig. S8, each exhibiting 1 to 3 major resonances. By isolation of these resonances individually, and taking care of zero-calibration according to the available center resonances in the data, the SI of each resonance has been calculated one by one. There are too many plots to put here, and instead only one plot containing all SI of all traces versus frequency of each resonance is shown in Fig. S9. This shows that even taking into consideration of uncertainties, the SI of each resonance could be as large as 4% in magnitude, a remarkably large deviation from zero.

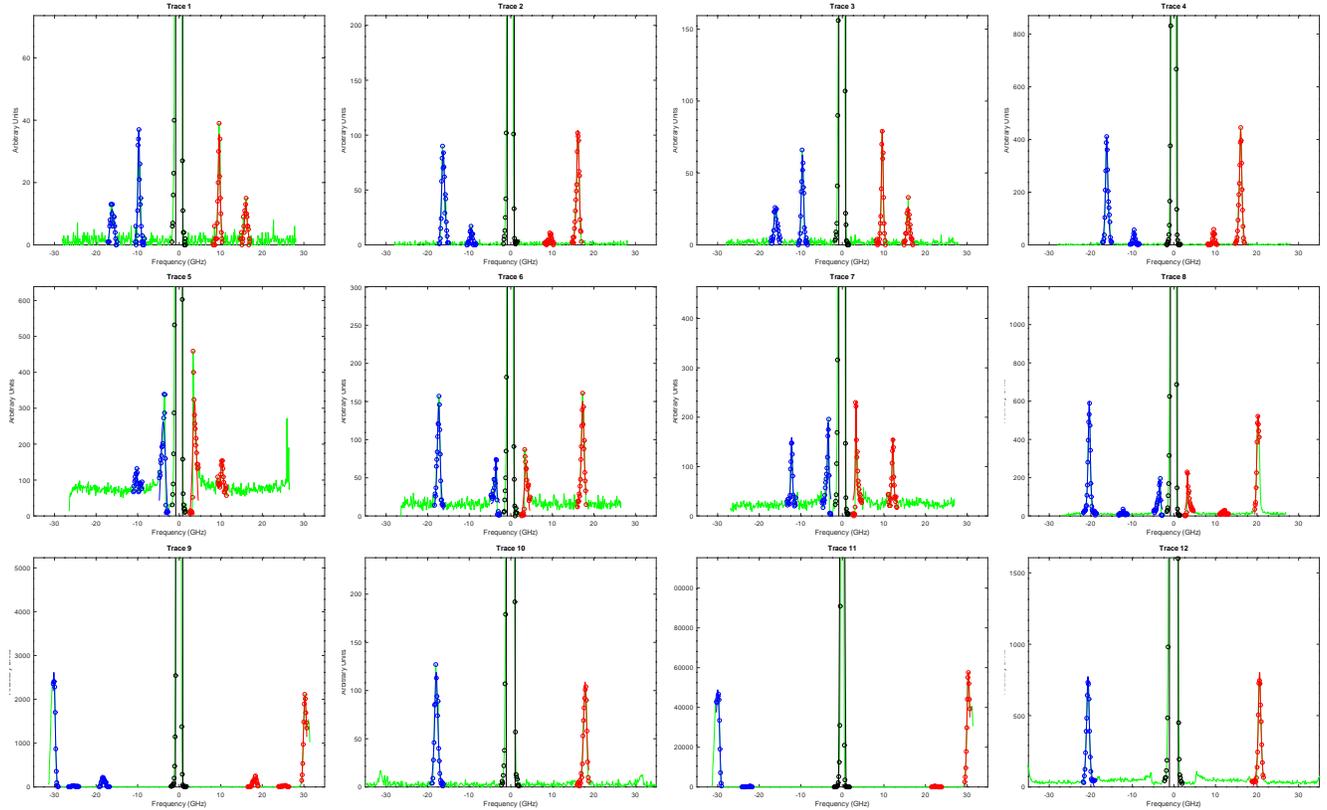

**Figure S8.** Hypersonic Brillouin scattering measurements for various polarizations and excitation conditions from raw measurement data [S49].

Finally, the last article in this category [S49] reports Brillouin scattering from phononic structures where accurate locations of side-bands are clearly marked throughout all figures therein, and is perfectly consistent with the positiveness of SI. The SI observed at various illumination angles for the two fabricated structures is always positive and reaches a value of 3.6%. The author was granted access to the raw measurement data, from which the figures S10,S11 were generated by making Lorentzian fits accurately to the relevant peaks.

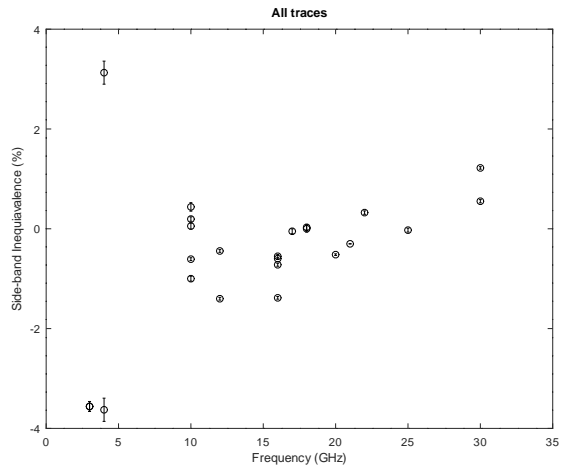

**Figure S9.** Collective SI points computed from raw measurement data of backward and forward Brillouin scattering of light for traces in Fig. S8, respectively with negative $\bar{\delta} < 0$ and positive $\bar{\delta} > 0$ SI for majority of points [S48].

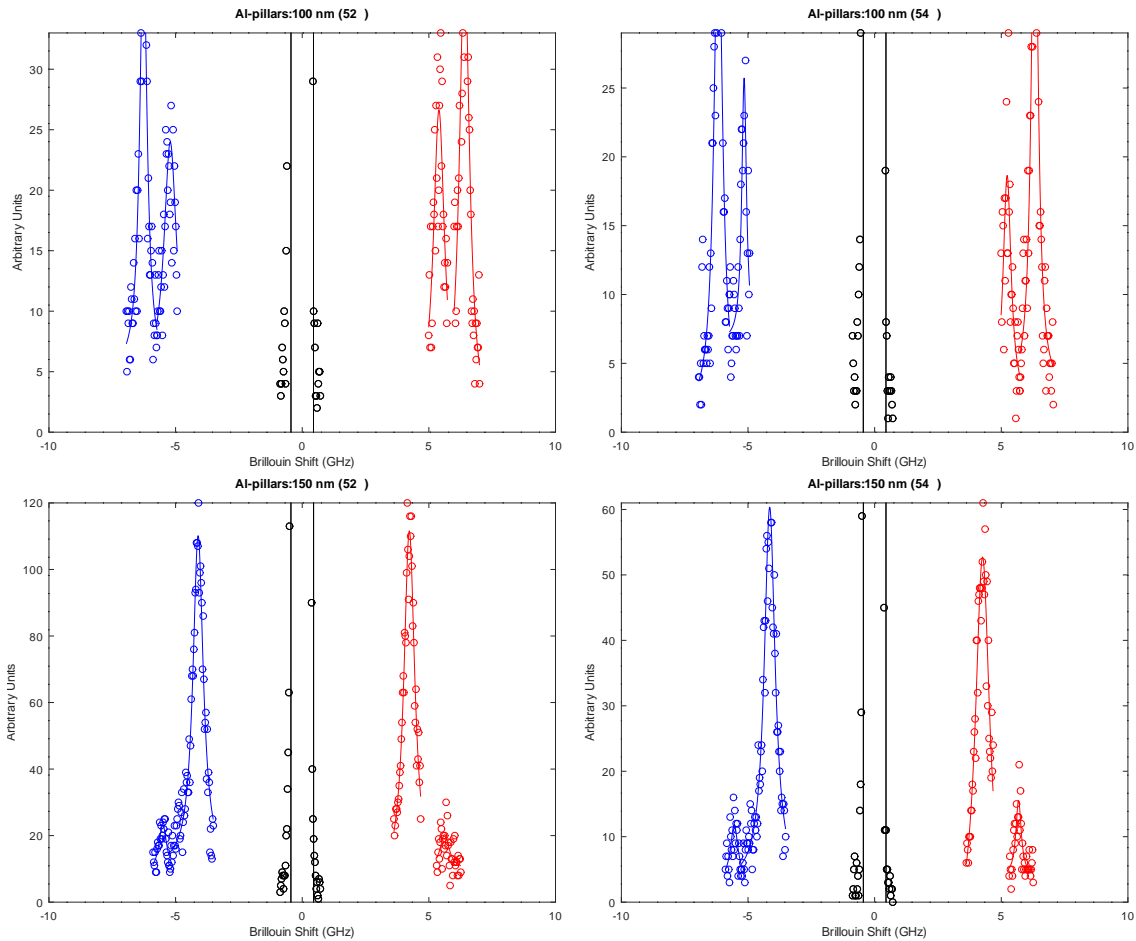

**Figure S10.** Lorentzian peaks for Stokes(red)/anti-Stokes(blue) pairs of Brillouin scattering measurements on Al nano pillars with different heights and illumination angles [S49].

## S5  Raman Scattering

Very few researchers actually happen to have noticed the existence of an anomaly in frequency asymmetry, and clearly made a mention of it. But in majority of published works, it has gone unnoticed. Among the bulky archives of available works over the

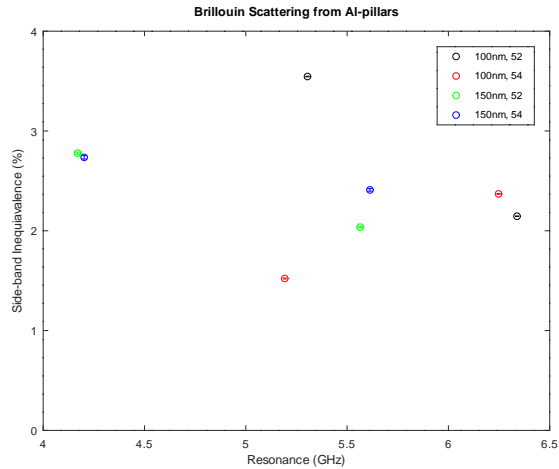

**Figure S11.** Side-band Inequivalences computed from raw measurement data [S49].

past decades, Raman scattering experiments provide the strongest SI of all.

However, Raman scattering is also the most complex one in the broad family of optomechanical types of interactions, where even for the simplest case of perfectly defect-free and ideal crystals, an integration over various phonon polarizations and momenta should be carried out. Hence, any non-ideal effects such as point or line defects and internal stress in piezo-electric materials could significantly alter the Raman response. For that reason, it is believed that measurements carried out on ultra-pure epitaxial cubic crystals such as Silicon (Si) are among the most reliable sources to verify the behavior of SI.

In crystalline solids, this normally reduces in effect to interaction with longitudinal optical phonons at van Hove singularities, which have nearly-identical frequencies and group velocities close to zero. For the exceptional case of Dirac materials such as graphite and graphene and metallic carbon nanotubes, the phonon dispersion together with the linear dispersion of photons can source another cause of frequency asymmetry, which normally goes in the opposite direction than that of SI. This effect however is optically linear, and does not scale with optical power.

Temperature contributions to Raman scattering are relatively significant, however, it mostly translates into enhanced optical spring effect which goes symmetrical in frequency shifts of red and blue side-bands. Since SI is all about the asymmetries, this contribution is automatically canceled out. Therefore, for graphite, the frequency asymmetry due to Anomalous Raman Phenomenon (ARP) is large enough to cause anti-Stokes shift exceed than that of Stokes resulting in $\tilde{\delta} < 0$.

In gapped crystalline dielectrics such as Si, SI is strong enough to cause Stokes shift being larger than anti-Stokes, and this is particularly easily observable for the SI in characteristic Raman line $520\text{cm}^{-1}$ of Si.

In summary, there appears to be two competing effects which collectively contribute to frequency asymmetry in Raman scattering:

1. Side-band Inequivalence (SI): which is frequency independent but optically nonlinear, and
2. Anomalous Raman Phenomenon (ARP): which is independent of optical power but frequency dependent.

Normally, SI and ARP cause asymmetry biased towards respectively red and blue. With the exception of Dirac materials, only for which ARP has been known to exist as early as 1998 [S50], in the rest of materials, ARP could be simply neglected. This implies that effectively ARP takes over for Dirac materials, which for other crystalline solids such as Si, SI does.

The authors of the first article referring to the ARP boldly mentioned that *ARP does not arise from the error of the instrument* [S50]. That was the first major step towards realizing that Stokes and anti-Stokes pairs are not necessarily symmetrical in frequency. Initial explanations were based on the formation of doublets [S51, S52], before it was found out that it was actually the slight deviation from linear photon dispersion across the Dirac point, which could be held responsible for ARP [S53–S56].

The expression for ARP was given as $\delta = -E_s(\partial \omega_s / \partial \varepsilon_L)$ with $E_s$ being the energy of Raman Stokes line, $\omega_s$ being the frequency of Raman shift, $\varepsilon_L$ being energy of light photon, and the expression within the parentheses being the dispersion of Raman peak determined from electronic energy band structure. Clearly, ARP is independent of optical power and therefore is optically linear. However, it could play measureable role in Raman spectroscopy of graphene [S57–S59] and even corrections to Raman thermometry and determination of temperature, if it were to be obtained from Stokes/anti-Stokes difference [S57].

Raman spectrometry of crystalline solids contain some of the most interesting results. In a recent article, Raman scattering of the zincblende semiconductor ZnTe at various temperatures has been carried out [S60] and the raw measurement data were made available to the author. Interestingly, the longitudial (LO) and transverse (TA) optical phonons contribute differently to

the Raman spectra and the second harmonic of 2LO can also be seen. Interestingly, for most of the temperature range, both of LO and 2LO modes follow the same behavior, within a factor of 2. This is in complete agreement with the formula for the SI of second-order side-bands in quantum optomechanics in the weakly nonlinear regime [S3], and thus another way to verify the higher-order operator algebra presented therein.

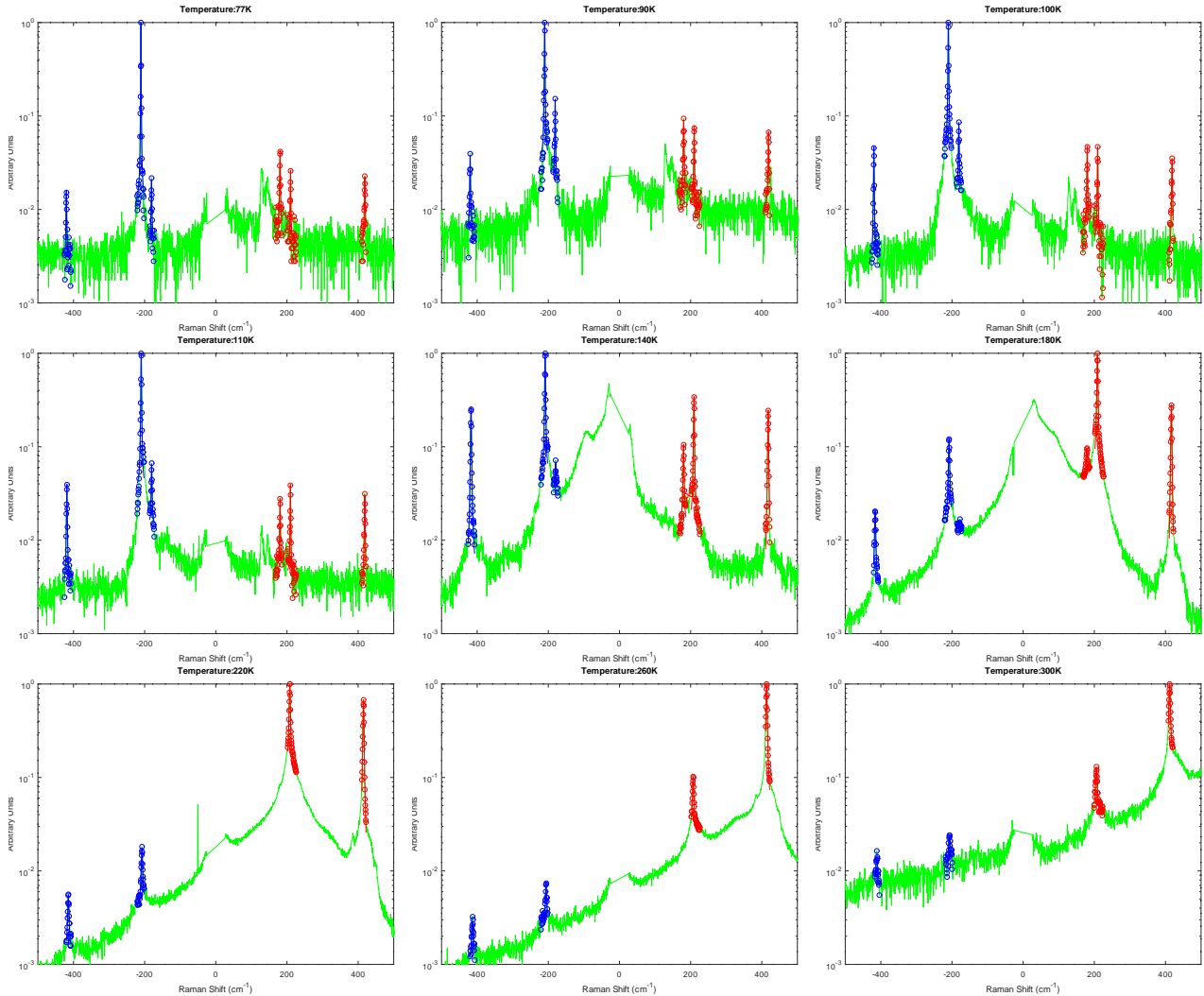

**Figure S12.** Lorentzian peaks for Stokes(red)/anti-Stokes(blue) pairs of Raman scattering measurements from ZnTe at various temperatures [S60].

In another study [S61], SI of three different materials, namely graphene on Si, MoS$_2$ on Si, and crystalline Si (with a 100nm native oxide) has been carried out at various temperatures. These are among the most extensive collection of measurements ever made available to the author in raw data format. Measurements were done using a 600gr/mm grating and for each trace both Stokes and anti-Stokes along with the central resonance were recorded at one shot. For those traces corresponding to the same material and the same temperature, trace-averages were made. Each trace exhibited a number of Raman lines which were extracted and after fitting Lorentzians and referencing to the frequency determined from the central Rayleigh peak, are shown in Fig. S14 at various temperatures. As it could be seen in Fig. S15, all SI values are positive and can be as large as 1% and decrease with Raman line frequency. Agreement to a $\tilde{\delta} \propto 1/\Omega^2$ fit from (S12) is excellent.

Similar measurements were done for MoS$_2$ on Si samples, and the SI values at various temperatures are shown for four individual Raman lines in Fig. S16. In general, a temperature variation from room temperature to 150°C does not seem to alter the SI values appreciably.

Among these sets of measurements[S61], the most interesting is the last set done for Si, which exhibits only two resonances at 303/cm and 520/cm, and the experiment was redone and traces recorded a few times to later provide for averaging. The SI values are shown in Fig. S17 for all available traces in two forms of versus temperature and versus optical power. In Fig. S18

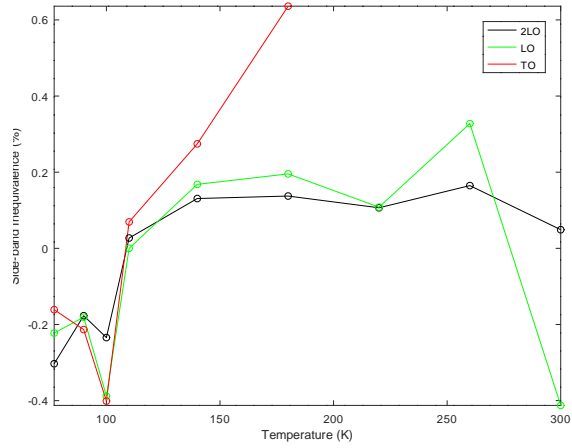

**Figure S13.** Side-band Inequivalences computed from raw measurement data processed in Fig. S12 [S60].

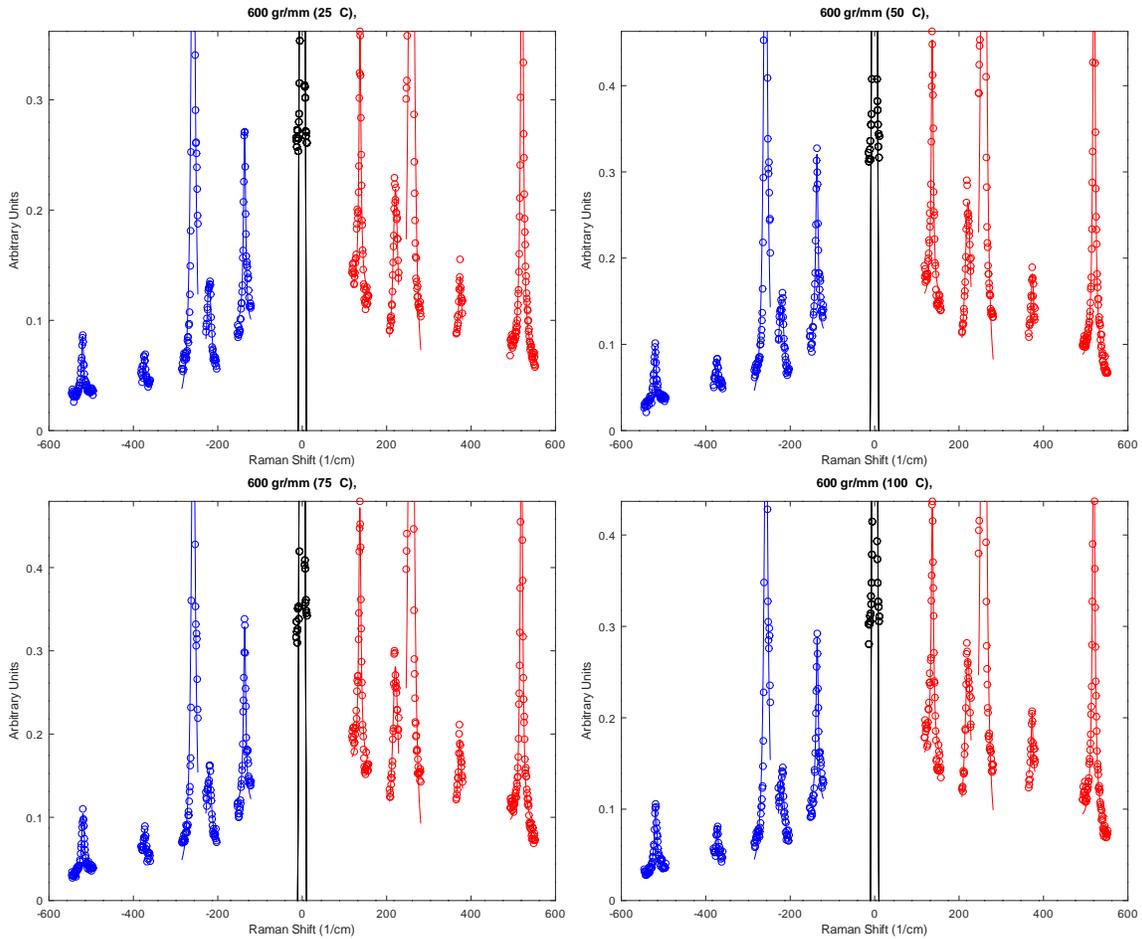

**Figure S14.** Lorentzian peaks for Stokes(red)/anti-Stokes(blue) pairs of Raman scattering measurements from graphene on Si substrate at various temperatures [S61].

after averaging over all traces at identical temperatures, as we learned from the previous two sets of traces.

This will make it possible to estimate a linear fit as $y = mx$ going through origin, where $(x_i, y_i)$ samples are available. This is so since in the weakly coupled limit, SI should zero at zero optical power, behaving roughly as $\delta \approx g_0^2 \bar{n} / \Omega$ [S3]. Then, the least-squares slope is simply obtained as $m = \sum(x_i y_i) / \sum x_i^2$. The ratio of obtained slopes for the two Raman lines is to be

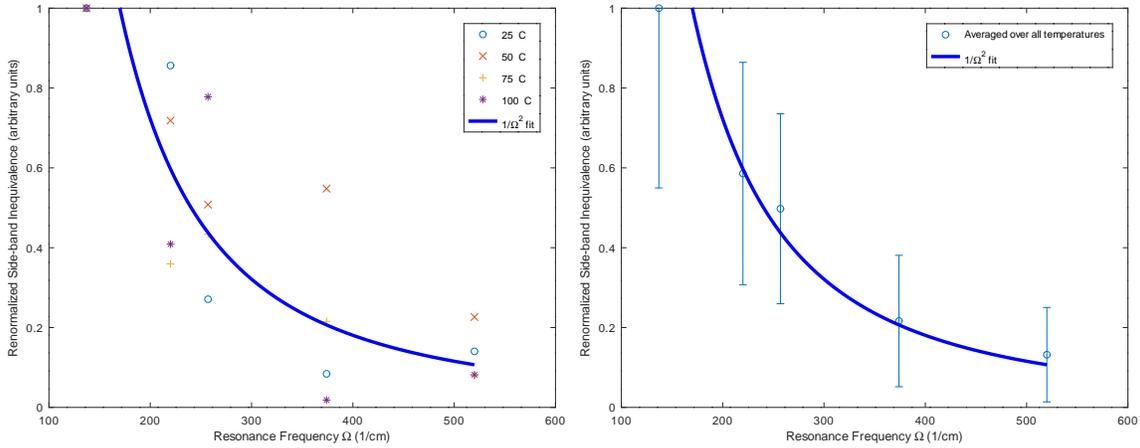

**Figure S15.** Side-band Inequivalences computed from raw measurement data processed in Fig. S14: SI for various Raman lines at different temperatures (left); SI for various Raman lines temperature-averaged [S61]. Note that SI values for each trace are renormalized to their maximum, to allow a simple and visible fit to (S12) which is in remarkable agreement.

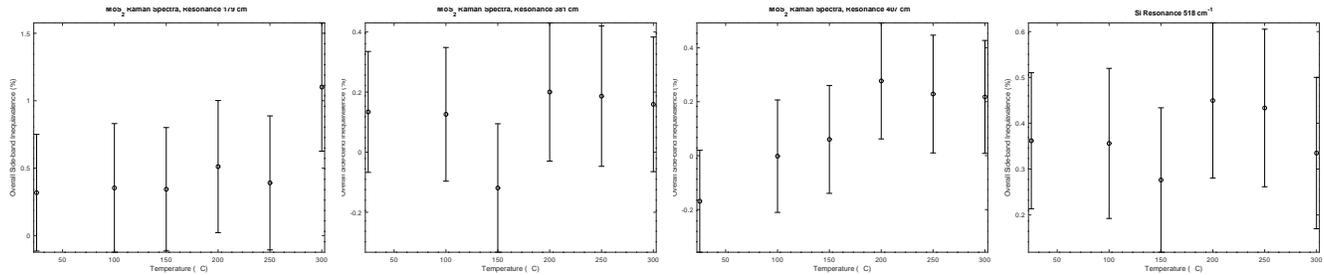

**Figure S16.** Variation of four identified Raman lines from MoS$_2$ on Si substrate with respect to substrate temperature [S61].

compared with the inverse ratio of squared frequencies and the results agree very well. Doing this will give

$$\frac{m_1}{m_2} = 0.33,$$

$$\left(\frac{\Omega_1}{\Omega_2}\right)^{-2} = 0.33,$$ (S13)

where subscripts 2 and 1 respectively correspond to the 303/cm and 520/cm lines.

Careful inspection of Raman scattering on two-dimensional Transition Metal Di-Chalcogenides (TMDCs) [S62, S63] exhibits significant SI for all materials at all resonances, which deserves a thorough and in-depth study.

A very recent article [S64] has reported Raman scattering study of two pervoskites CsPbBr$_3$ and MAPbBr$_3$ at various temperatures. Pervoskites are known to be very complex materials with peculiar optoelectronic and chemical properties, which are not fully understood, yet exhibit numerous Raman lines. The raw measurement data were made available to the author from which SI values of identified Raman peaks were extracted. After analysis and averaging over temperatures, the overall SI values have been found and plotted in Fig. S19. Calculated SI numbers are typically large in magnitude up to 3% and do not follow a well understood pattern. Also, Raman measurements on NbSe$_2$ [S65] for the 11-layer and bulk samples respectively show SI values of 0.46% and 4.56%, the latter being remarkably large.

In a valuable book [S66] which also contains the demonstration version of OPUS [S67] software for plotting Raman spectra, a database of measurement is supplied out of which traces of four materials contain both Stokes and anti-Stokes side-bands. Typically, these reference measurements are averaged over hundreds of Raman scattering to remove background noise and improve the visibility of resonances. To a great extent, these measurements seem reliable and furthermore, contain the central Rayleigh resonance which allows accurate zero centering. These four mentioned substances are Ceramide, heavy Ethanol, Stearic Acid, and Sulphur. With the exception of first resonance of Stearic Acid, all these exhibit remarkably strong and positive SI, as shown in Fig. S20.

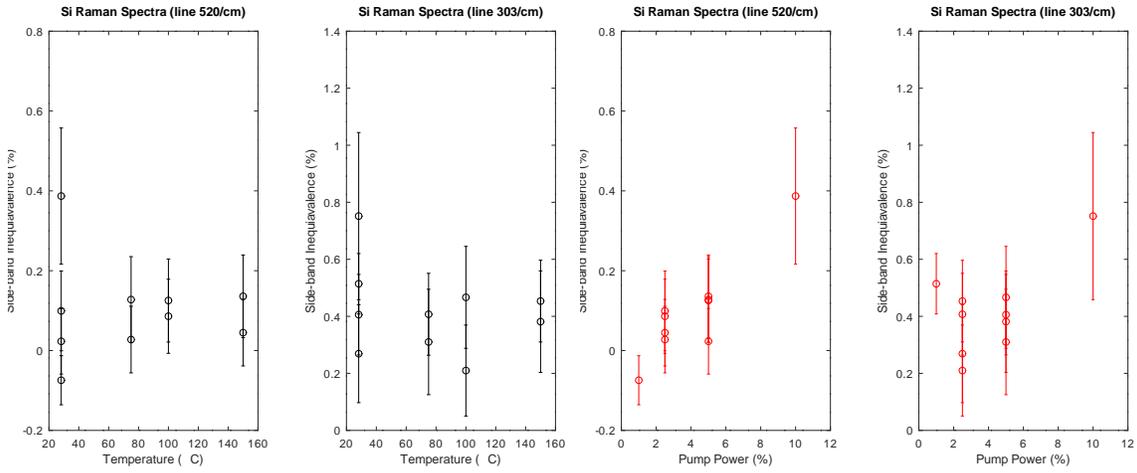

**Figure S17.** Variation of SI for two characteristic lines of Si versus temperature (left in black) and versus optical power (right in red) for available traces [S61].

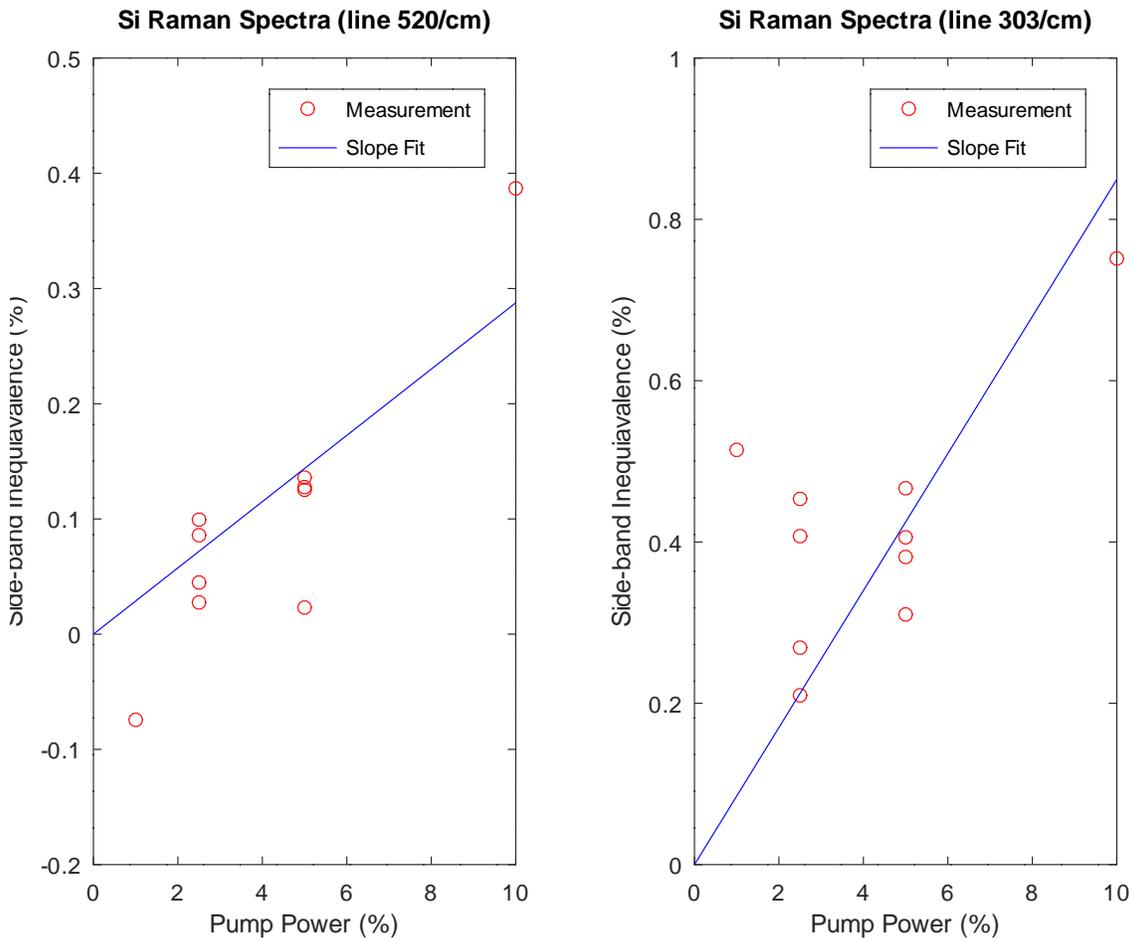

**Figure S18.** Variation of SI versus optical power for the two major lines at 303/cm and 520/cm [S61].

The Coherent company has developed a new technology for precision remote Raman analysis of various materials, many examples of which are publicized in their newest brochure [S68]. It is very interesting to take a look at the calculated SI values from their available data. All these very different substances, (1) Carbamazepine (a common prescription for epilepsy and other

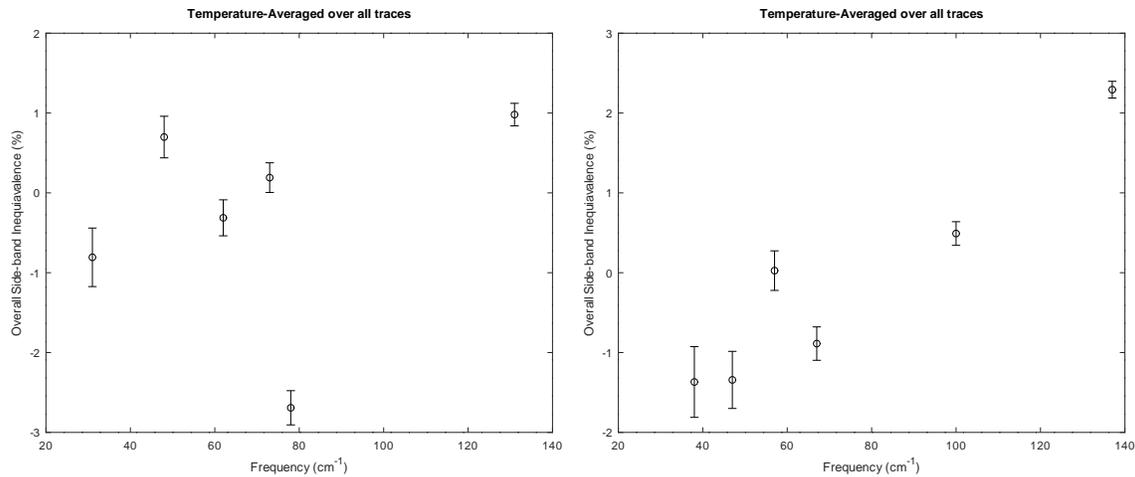

**Figure S19.** Variation of SI after temperature averaging of identified Raman lines for two pervoskites CsPbBr$_3$ and MAPbBr$_3$ [S64] from raw measurement data.

neurologic disorders), (2) Carbamazepine hydrate, (3) Crystallized Caffeine (the extract and essential ingredient of coffee), and (4) Sulphur in three different configurations ($\alpha$: crystalline; $\beta$: amorphous; $\lambda$: liquid) exhibit large SI numbers up for many Raman lines to a few percent. In all four studied cases, SI is positive and significantly larger than zero.

In the end, Crystallized Coffee exhibits five Raman lines at 42.6/cm, 61.5/cm, 105.4/cm, 122.6/cm, and 180.1/cm, with SI values respectively equal to 2.06%, 1.43%, 0.833%, 0.717%, and 0.488%. Surprisingly or not, these very well fit to (S12) with a total optomechanical decay rate of $\gamma = 2\pi \times 25 \text{cm}^{-1}$, reminding Alfréd Rényi and legendary Paul Erdös who believed [S69] "*A mathematician is a machine for turning coffee into theorems.*"

## Acknowledgement

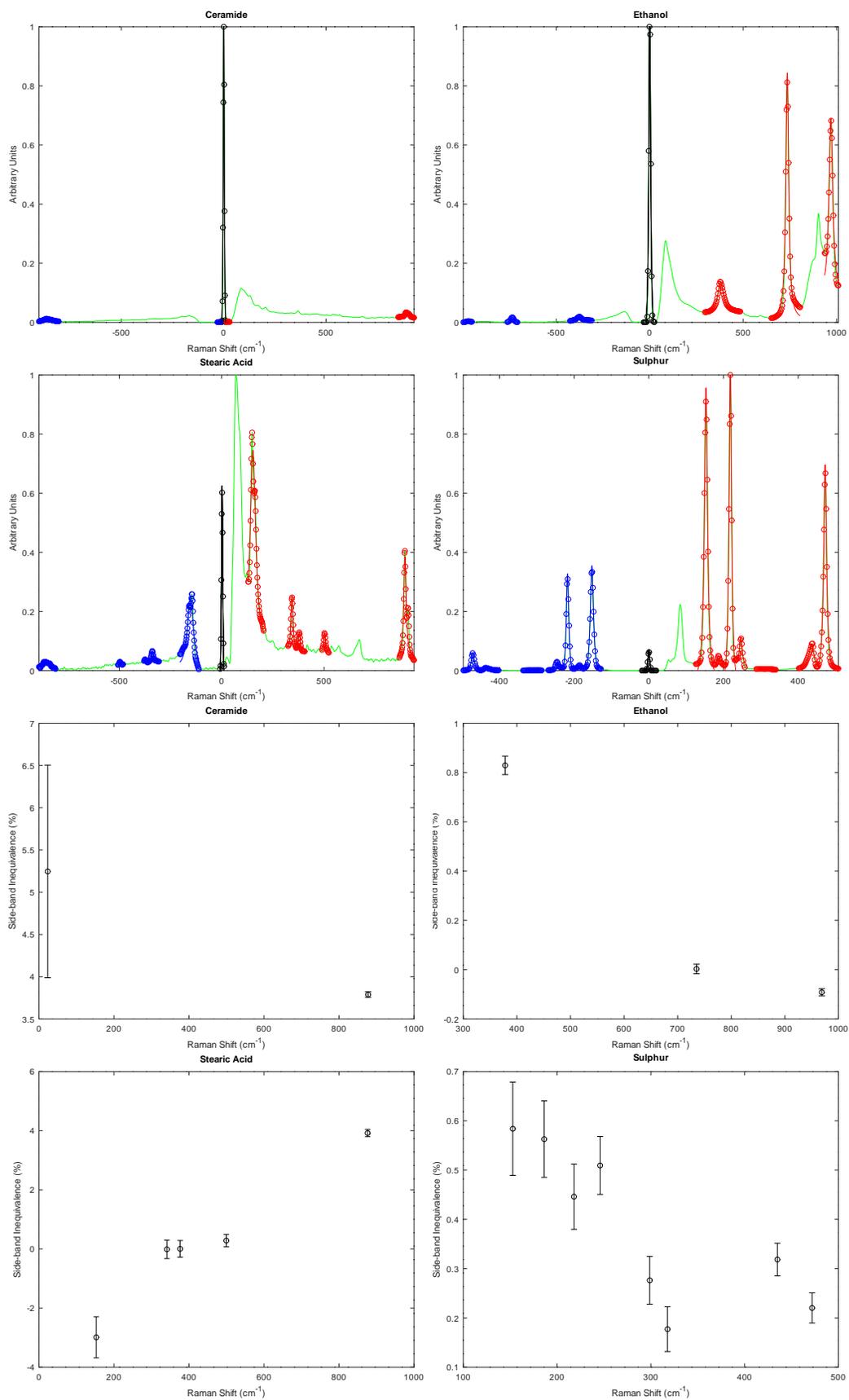

**Figure S20.** Raman spectra (top rows) and corresponding SI values (bottom rows) from raw measurement data [S65] for Ceramide, Ethanol, Stearic Acid, and Sulphur.

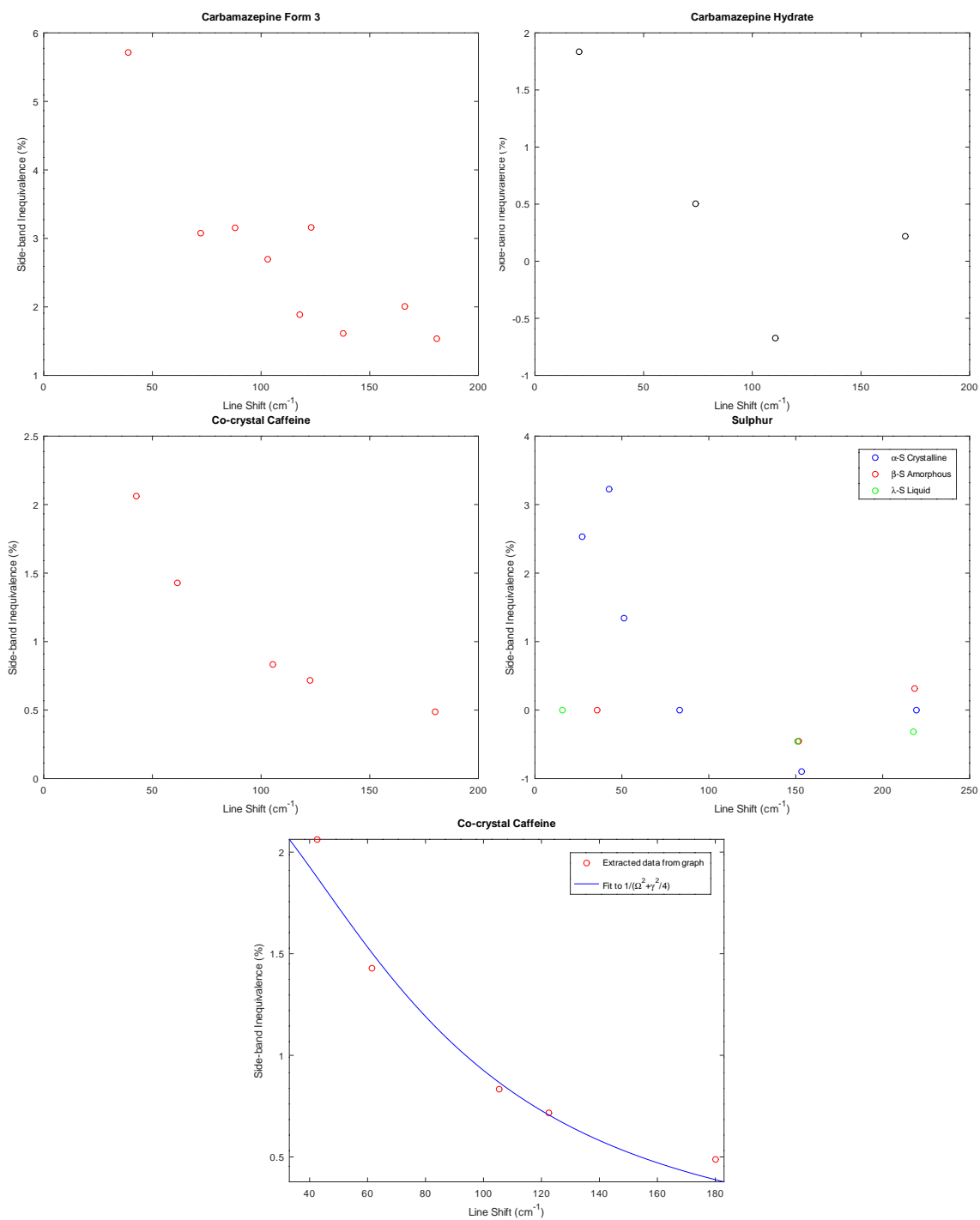

**Figure S21.** Calculated SI values extracted for four very different materials using a new THz/Raman scattering technology: (top-left) Carbamazepine, a common medication for epilepsy and neurologic disorders; (top-right) Carbamazepine hydrate; (middle-left) Crystallized Caffeine, the extract of beloved hot drink coffee (middle-right) Sulphur in three different configurations ($\alpha$: crystalline; $\beta$: amorphous; $\lambda$: liquid); (bottom) SI for Crystallized Caffeine extracted from measured graph, with fitting to (S12) with $\gamma = 157.1\text{cm}^{-1}$ [S68].



\end{document}